\documentstyle[12pt,epsf]{article}
\def\siml{{\ \lower-1.2pt\vbox{\hbox{\rlap{$<$}\lower6pt\vbox{\hbox{$\sim$}}}}\ }}

\def\als{\alpha_{\rm s}}
\def\al{\alpha}
\def\lQ{\Lambda_{\rm QCD}}

\newcommand{\Appendix}[1]%
    {%
     \section{#1}%
      }

\begin{document}
\begin{titlepage}
\begin{flushright}
\tt{CERN-TH/99-199\\ HEPHY-PUB 716/99\\ UB-ECM-PF 99/13\\ UWThPh-1999-34}
\end{flushright}

\vspace{1cm}
\begin{center}
\begin{Large}
{\bf Potential NRQCD: an effective theory for heavy quarkonium}\\[2cm]
\end{Large} 
{\large Nora Brambilla}\footnote{brambill@doppler.thp.univie.ac.at}\\
{\it Institut f\"ur Theoretische Physik, Boltzmanngasse 5, A-1090 Vienna,
  Austria}\\[0.6cm] 
{\large Antonio Pineda}\footnote{antonio.pineda@cern.ch}\\
{\it Theory Division CERN, 1211 Geneva 23, Switzerland}\\[0.6cm]
{\large Joan Soto}\footnote{soto@ecm.ub.es}\\
{\it Dept. d'Estructura i Constituents de la Mat\`eria and IFAE, U. Barcelona \\ 
     Diagonal 647, E-08028 Barcelona, Catalonia, Spain}\\[0.6cm]
{\large Antonio Vairo}\footnote{vairo@hephy.oeaw.ac.at}\\
{\it Institut  f\"ur Hochenergiephysik, \"Osterr. Akademie d. Wissenschaften\\
     Nikolsdorfergasse 18, A-1050 Vienna, Austria}
\end{center}

\vspace{1cm}

\begin{abstract}
\noindent
Within an effective field theory framework we study heavy-quark--anti\-quark
systems with a typical distance between the heavy quark and the antiquark smaller 
than $1/\Lambda_{\rm QCD}$. A suitable definition of the potential is given within this framework, 
while non-potential (retardation) effects are taken into account in a systematic way. 
We explore different physical systems. Model-independent results on the
short distance behavior of the energies of the gluonic excitations between
static quarks are obtained. Finally, we show how infrared renormalons affecting the static potential 
get cancelled in the effective theory. 
\vspace{5mm} \\
PACS numbers: 12.38.Aw, 12.38.Bx, 12.39.Hg 
\end{abstract}

\end{titlepage}
\vfill
\setcounter{footnote}{0} 

\section{Introduction}
It is an experimental evidence that for heavy-quark bound states
the orbital splittings are smaller than the quark mass $m$.
This suggests that all the other dynamical 
scales of these systems 
are smaller than $m$. 
Consistently with this fact, the quark velocity $v$ in these 
systems 
is believed to
be  a small quantity, $v \ll 1$. Therefore, 
a non-relativistic (NR) picture  holds \cite{old}. 
This
produces a hierarchy of scales: $m\gg mv \gg mv^2  $ in the system 
\cite{NRQCD}. 
The inverse of the soft scale, $mv$, gives the size of the 
bound state, while the inverse of the ultrasoft scale, $mv^2$, 
gives the typical time scale. However, in QCD  another physically relevant scale 
has to be considered, namely 
the scale at which nonperturbative effects become important, which we will generically denote by 
$\lQ$.

The study of the heavy quarkonium properties was initiated long ago, using potential models.
A large variety of them exists in the literature and they have been on the
whole phenomenologically quite successful (see \cite{potmod} for reviews).
However, their connection with the QCD parameters is hidden, the scale at which they are defined  is not
clear, and they cannot be systematically improved. 
In spite of this, great progress has been achieved 
over the years by relating the potentials appearing in these models to some static Wilson loop operators \cite{pot1,pot2,pot3,
pot4}. 
This formulation is particularly suitable in QCD because it enables a direct lattice calculation 
and/or  an analytic  calculation in a QCD vacuum model \cite{mod}.

However, there still remains the 
question of to which extent this pure-potential picture is 
accurate \cite{Appelquist,Brown,pNRQCD,BPSV,pot4}. 
In order to answer this question it seems necessary to develop a formalism where 
the error produced by using a Schr\"odinger 
equation with a potential obtained from 
QCD (e.g. by lattice simulations \cite{bali}) instead of doing 
the computation in full QCD (e.g. by NRQCD lattice simulations \cite{lepagelattice})
can be made quantitative \cite{lepage}. In this work 
we provide the first steps towards this goal.  

\medskip

In QED, non-potential effects (sometimes noted in the literature as retardation
effects) arise as $O(\alpha^3)$ corrections to the energy levels (Lamb shift). 
They are due to (ultrasoft) photons with an energy of $O(mv^2)$. Their  existence is signalled in QED  
by an infrared (IR) divergence in the $O(1/m^2)$ potential. Hence, potential models for QED are only accurate up to 
$O(\alpha^2)$ corrections. In spite of that, the static potential 
is well defined at any order of perturbation theory and coincides with the Coulomb potential. 
In perturbative QCD the situation may be regarded  as similar or very different from QED, depending 
on the object we are interested in. On the one hand, non-potential effects in perturbative 
quarkonium bound states appear as $O(\alpha_{\rm s}^3)$ corrections as in QED, and also due to 
ultrasoft gluons with energy of $O(mv^2)$ ($v\sim \alpha_{\rm s} $). On the other hand, the static potential suffers 
from IR divergences in perturbative QCD \cite{Appelquist}, and hence, in this respect, the situation 
is completely different from QED. 
 In any case, potential models for perturbative QCD are only accurate up to $O(\alpha_{\rm s}^2)$ corrections. 
The complete $O(\alpha_{\rm s}^2)$ corrections to the perturbative static potential 
have now been calculated \cite{twoloop}, and very recently we have presented 
the leading log ($\mu$-dependent) contribution to the $O(\alpha_{\rm s}^3)$
correction \cite{BPSV}. The next perturbative improvement in the determination
of the bottom quark mass from the $\Upsilon (1S)$ system
\cite{Yndurain,Beneke1}, as well as the expected precision at the  Next Linear Collider for top pair production near threshold,
 are of this order \cite{Beneke}. Hence a systematic treatment of the 
potential and non-potential effects is also important in perturbative QCD for physical 
applications. Here, we shall further elaborate on this point. 

\medskip

A more complicated issue is how nonperturbative effects in QCD influence the above facts. 
Obviously, it will depend on the relative size of $\lQ$ with respect to the other dynamical scales.
To our knowledge, only in the situation where $m\gg mv \gg mv^2\gg \lQ$, can statements be made neatly. 
In this situation, which holds for the lowest-lying states of very heavy quarkonium, the leading nonperturbative corrections can be parameterized by local 
condensates \cite{Voloshin}, which have a non-potential origin. Our goal here
is to discuss other 
situations where neat statements can be made, restricting ourselves to the situation $mv\gg\lQ$ throughout this paper, 
but allowing for any relative size between $\lQ$ and $mv^2$.

\medskip

It is our general aim to take advantage of the remarkable progress that has been made in 
effective field theories for heavy quarks during the last years
\cite{NRQCD,Neubert,Labelle,Manohar,GR,pNRQCD,Beneke2,Match, pos,pot4,BPSV} to address the various issues mentioned above in a solid QCD-based framework. Since the hard ($m$), soft ($mv$) 
and ultrasoft ($mv^2$) scales are widely separated, two EFTs can be introduced by sequentially 
integrating out $m$ and $mv$. Upon integrating out the hard scale $m$, non-relativistic QCD (NRQCD) 
is obtained \cite{NRQCD}. Proceeding one step further and integrating out the soft scale $mv$, 
potential NRQCD (pNRQCD) is obtained \cite{pNRQCD}.
 This sequence of EFTs has the 
advantage that it allows disentangling of perturbative contributions from
nonperturbative ones to a large extent, which is of the utmost
importance in QCD.

\medskip

In this work we shall further study pNRQCD, as we define it in detail in section 2. 
This effective field theory contains a singlet and an octet field, which depend 
on the relative (${\bf r}$) and centre-of-mass (${\bf R}$) quark--antiquark
coordinates and on time; 
potential terms, which depend on the relative coordinate and the momentum (and spin);
ultrasoft gluon fields, which depend on the centre-of-mass coordinate and on time. Since pNRQCD 
has potential terms, it embraces potential models. Since in addition it still has ultrasoft gluons 
as dynamical degrees of freedom, it is able to describe non-potential effects.
Indeed, the QED version of it, namely pNRQED, has been shown to correctly reproduce the 
non-potential effects that arise as $O(\alpha^3)$ corrections to the binding energies of hydrogen-like 
atoms \cite{Lamb} and positronium \cite{pos}. Moreover, pNRQCD provides a new interpretation 
of the potentials that appear in the Schr\"odinger equation in terms of a modern effective 
field theory language. The potentials are nothing but {\bf r}-dependent matching coefficients, which appear 
after integrating out scales of the order of the relative momentum in the bound state ($\sim mv\sim 1/r$) or any remaining dynamical scale above $mv^2$. 
Being matching coefficients, the potentials generically depend on a 
subtraction point $\mu$. This is related to the IR divergences found in \cite{Appelquist}. 
The subtraction point $\mu$ may be regarded as the IR cut-off, which is
necessary to define 
the QCD static potential to any finite order of perturbation theory beyond two loops. 
In pNRQCD this $\mu$-dependence is natural and should not cause any concern. Indeed, below the scale 
$mv$ there are still dynamical gluons, which are incorporated in the pNRQCD 
Lagrangian. No gluons of energy higher than $\mu$ are allowed. Hence the dynamics of 
gluons in pNRQCD is cut-off by $\mu$, and this $\mu$ dependence will cancel the explicit $\mu$
dependence of the potential terms when calculating a physical
quantity\footnote{Some potentials may also have an extra IR scale
  dependence, inherited from the NRQCD matching coefficients, which cancels
  with the UV divergences in pNRQCD arising from
  quantum mechanical perturbation theory and not from the US gluons. This
  particular scale dependence would cancel against a scale of $O(mv)$, the
  typical relative momentum in the bound state, and not against the US scale
  of $O(mv^2)$. Since these effects appear at $O(1/m^2)$ and we shall
  deal with $O(1/m^0)$ potentials only, we will ignore this possibility in the
  present work.}. Once the interpretation 
of the potentials as matching coefficients is taken seriously, one may look for the 
running of the potentials. We have calculated the leading contributions to the running of the 
singlet, octet and mixed matching potentials (to be defined later) in perturbation theory.

We enforce the equivalence of pNRQCD to NRQCD, and hence to QCD, to the desired order in the multipole 
expansion by requiring the Green functions of both effective theories to be equal (matching).
The matching between NRQCD and pNRQCD can be done order by order in $1/m$.
This is due to the fact that we are integrating  out, in addition to gluons of momentum $\sim mv$, 
quarks with energies $k^0$ of the order of $mv$, which are always larger than the kinetic 
energy ${\bf k}^2/2m$ of the heavy quark for ${\bf k}\sim mv$, so that the latter can 
be expanded in the quark propagator (this will be made more precise in sections 2 and 3).  
If we denote by $\Lambda_{mp}$ any scale smaller than $mv$, at each order in $1/m$, the calculation 
in NRQCD contains, in general, arbitrary powers of $\Lambda_{mp} / mv$, while in  pNRQCD the counting 
in $\Lambda_{mp} / mv$ is explicit and we can decide at which order in $\Lambda_{mp} / mv$ 
we want to carry out the matching. This is nothing but the multipole expansion.
Moreover, if we restrict ourselves to the situation $mv \gg \lQ$, we can, in addition, 
do the matching order by order in perturbation theory. Then
at the leading 
order in $1/m$ and in the multipole expansion, the pNRQCD Lagrangian only 
depends on the perturbative singlet and octet potentials. The perturbative 
singlet potential at this order may be obtained from the standard calculation of the static Wilson loop. 
The perturbative octet potential may also be obtained from the static Wilson loop with suitable 
operator insertions at the end-point strings. At the next-to-leading order in
the multipole expansion, the pNRQCD Lagrangian contains in addition  mixed 
perturbative potentials, which  involve interactions with gluons of energy smaller than $mv$. 
Furthermore, at the same order, the perturbative singlet potential is no
longer given only by the static Wilson loop, nor is the perturbative octet
potential. This is due to the fact 
that the static Wilson loop contains contributions from scales below $mv$, which must be subtracted 
in order to have the perturbative static potential properly defined as a matching coefficient, 
namely as an object that has contributions only from scales of $O(mv)$. 
The static potential thus defined is the relevant quantity  to be used 
in the Schr\"odinger equation when the next relevant scale is $mv^2$.

In the situation where $\lQ\gg mv^2$, there is still a  dynamical 
scale above $mv^2$. In this case, pNRQCD is not yet the suitable EFT at the
ultrasoft scale $\sim mv^2$; 
neither is the singlet potential of pNRQCD the suitable potential to be used in the Schr\"odinger equation. 
In order to obtain the relevant Lagrangian which describes the ultrasoft degrees of freedom ($\sim mv^2$), 
and hence the relevant potential to be used in the Schr\"odinger equation, 
we still have to proceed one step further and integrate out scales $\sim \lQ$. 
Of course, this can no longer  be carried out perturbatively in $\alpha_{\rm s}$. 
The scales larger than $mv^2$ (but smaller than $mv$) give rise to nonperturbative contributions 
to the new potential. We will discuss this situation in section \ref{secl2}.

\medskip

We distribute the paper as follows. In section \ref{sectheory} we review
pNRQCD in detail. In section \ref{secmatch} we discuss in detail the matching between NRQCD and pNRQCD 
at leading order in $1/m$ and at next-to-leading order in the multipole expansion. We also  
present the leading contributions to the running of the potentials in perturbation theory. 
In sections \ref{secl1} and \ref{secl2} we discuss the various physical
situations that arise depending on the relative size of $\lQ$ with
respect to $mv^2$. In section  \ref{hybrids} some 
model-independent results on the short-distance limit of the gluonic
excitations between static quarks are obtained. Section \ref{seccon} is
devoted to an outlook and conclusions. In \ref{matgenrem} we discuss the general
matching structure, while in \ref{secrenormalon} we show how IR renormalons
affecting the static potential get cancelled in the effective theory.
 
\section{Theoretical Framework}
\label{sectheory}

\subsection{NRQCD}
Integrating out from QCD the hard scale ($\sim m$) while considering almost on-shell heavy quarks and antiquarks 
produces NRQCD \cite{NRQCD}.  As noticed in \cite{Manohar,Match}, at this stage it coincides with 
the Heavy Quark Effective Theory (HQET). It describes degrees of freedom (quarks and gluons) with 
energy and momentum less than a certain cut-off, which is much smaller than $m$ and much larger 
than any remaining physical scale. The NRQCD Lagrangian is written as a power expansion in  $1/m$. 
The maximum size of each term may be obtained by assigning the next relevant scale  
(usually $mv$ in NRQCD and $\lQ$ in HQET) to any dimensional operator.
For the purposes of this work it is sufficient to use the NRQCD Lagrangian at the lowest order in $1/m$:
\begin{equation}
{\cal L}_{\rm NRQCD} \! = \! \psi^\dagger \, i D^0 \, \psi + \chi^\dagger \, i D^0 
\, \chi - {1\over 4} F_{\mu \nu}^{a} F^{\mu \nu \, a},
\label{lagnrqcd}
\end{equation}
where $\psi$ is the Pauli spinor field that annihilates the fermion and $\chi$
is the Pauli spinor field that creates the antifermion, $i D^0=i\partial_0 -gA^0$.

\subsection{pNRQCD: the degrees of freedom}
Integrating out the soft scale, $mv$, in  (\ref{lagnrqcd}) produces  pNRQCD \cite{pNRQCD}. 
The relevant degrees of freedom of pNRQCD may depend in general on the nonperturbative features of NRQCD. 
In this work we will assume that there exists a matching scale $\mu$ such that $mv\gg\mu\gg mv^2,\, \lQ$, where 
a perturbative picture still  holds. 

Strictly speaking pNRQCD has two ultraviolet (UV) cut-offs $\Lambda_1$ and $\Lambda_2$. 
The former fulfils the relation  $ mv^2,\lQ$  $\ll \Lambda_1 \ll$  $mv$ and is the cut-off of the
energy of the quarks and of the energy and the momentum of the gluons (it corresponds to the $\mu$ above), 
whereas the latter fulfils $mv \ll \Lambda_2 \ll m$ and is the cut-off of the relative
momentum of the quark--antiquark system, ${\bf p}$.\footnote{Notice
that although, for simplicity, we usually refer to the matching between
NRQCD and pNRQCD
as integrating out the soft scale, it should be clear from this
paragraph that the relative
momentum of the quarks in pNRQCD is still soft, and hence it has not been
integrated out.} In principle, we have some
freedom in choosing the relative importance between $\Lambda_1$ and
$\Lambda_2$. Although it is not necessary for the calculations of this work,
it is convenient to have in mind the choice $\Lambda_2^2/m \ll  \Lambda_1$,
which guarantees that the UV behaviour of the $Q$-$\bar Q$ propagator in pNRQCD
is that of the static one when virtual energies are present\footnote{Of course, this is not so for fixed energy and virtual relative
  momenta. In this case the UV behaviour is dominated by the relative momentum kinetic
term.}. 

If we denote any scale below $\Lambda_1$, i.e. $mv^2,\lQ$, ..., with  $\Lambda_{mp}$, 
we are in a position to enumerate the effective degrees of freedom of pNRQCD.
These are: $Q$-$\bar Q$ states with energy of $O(\Lambda_{mp})$ and relative momentum not larger 
than the soft scale; gluons with energy and momentum of $O(\Lambda_{mp})$. 
Let us define the centre-of-mass coordinate of the $Q$-$\bar Q$ system 
${\bf R} \equiv ({\bf x}_1+{\bf x}_2)/2$ and the relative coordinate ${\bf r} \equiv {\bf x}_1-{\bf x}_2$. 
A $Q$-$\bar Q$ state can be decomposed into a singlet state $S({\bf R},{\bf r},t)$ and an octet 
state $O({\bf R},{\bf r},t)$, in relation to colour gauge transformation 
with respect to the centre-of-mass coordinate. We notice that in QED only the analogous to the singlet appears. 
The gauge fields are evaluated in ${\bf R}$ and $t$, i.e.  $A_\mu = A_\mu({\bf R},t)$: 
they do not depend on ${\bf r}$. This is due to the fact that, since the typical size  ${\bf r}$ 
is the inverse of the soft scale, gluon fields are multipole expanded with respect to this variable.

\subsection{Power counting}
In order to discuss the general structure of the pNRQCD Lagrangian, let us consider  in more detail 
the different scales involved into the problem 
$$
m\,, \quad p\,, \quad {1 \over r} \, ,\quad \Lambda_{mp}\,.
$$
The variables $r$ and $m$ will explicitly appear in the pNRQCD Lagrangian, since they 
correspond to scales that have been integrated out. Although $p$ and $1 / r$ are of the same size 
in the physical system, we will keep them as independent. This will facilitate the counting rules 
used to build the most general Lagrangian. 

With the above objects, the following small dimensionless quantities will appear 
\begin{equation}
{p \over m},\quad {1 \over r m}, \quad \Lambda_{mp}r \quad \ll 1.
\label{uneq1}
\end{equation}
Note that $1/p r$ or $\Lambda_{mp} /p$ are not allowed since $p$ has to appear in an analytic way 
in the Lagrangian (this scale has not been integrated out). 
The last inequality in Eq. (\ref{uneq1}) tells us that $r$ can be considered
to be small 
with respect to the remaining dynamical lengths in the system. As a consequence the 
gluon fields can be systematically expanded in $r$ (multipole expansion).

Therefore, the pNRQCD Lagrangian can be written as an expansion in $1/m$ 
(from the first two inequalities of Eq. (\ref{uneq1})), 
and as an expansion in $r$ (the so-called multipole expansion, corresponding to the 
third inequality of Eq. (\ref{uneq1})). As a typical feature of an 
effective theory, the non-analytic behaviour in $r$ is encoded in the matching coefficients.

\subsection{pNRQCD: the Lagrangian}
The most general pNRQCD Lagrangian density that can be constructed with the fields 
introduced above and that is compatible with the symmetries of NRQCD is given, at leading order 
in $1/m$ and at  $O(r)$ in the multipole expansion, by:
\begin{eqnarray}
& & {\cal L}_{\rm pNRQCD} =
{\rm Tr} \,\Biggl\{ {\rm S}^\dagger \left( i\partial_0 - V_s(r) + \dots  \right) {\rm S} 
+ {\rm O}^\dagger \left( iD_0 - V_o(r) + \dots  \right) {\rm O} \Biggr\}
\nonumber\\
\nonumber
& &\qquad + g V_A ( r) {\rm Tr} \left\{  {\rm O}^\dagger {\bf r} \cdot {\bf E} \,{\rm S}
+ {\rm S}^\dagger {\bf r} \cdot {\bf E} \,{\rm O} \right\} 
+ g {V_B (r) \over 2} {\rm Tr} \left\{  {\rm O}^\dagger {\bf r} \cdot {\bf E} \, {\rm O} 
+ {\rm O}^\dagger {\rm O} {\bf r} \cdot {\bf E}  \right\}  
\\
& &\qquad- {1\over 4} F_{\mu \nu}^{a} F^{\mu \nu \, a}\,.
\label{pnrqcd0}
\end{eqnarray}
All the gauge fields in Eq. (\ref {pnrqcd0}) are evaluated 
in ${\bf R}$ and $t$, in particular $F^{\mu \nu \, a} \equiv F^{\mu \nu \, a}
({\bf R},t)$ and $iD_0 {\rm O} \equiv i \partial_0 {\rm O} - g [A_0({\bf R},t),{\rm O}]$. 

\begin{figure}[htb]
\makebox[0.5cm]{\phantom b}
\epsfxsize=3truecm \epsfbox{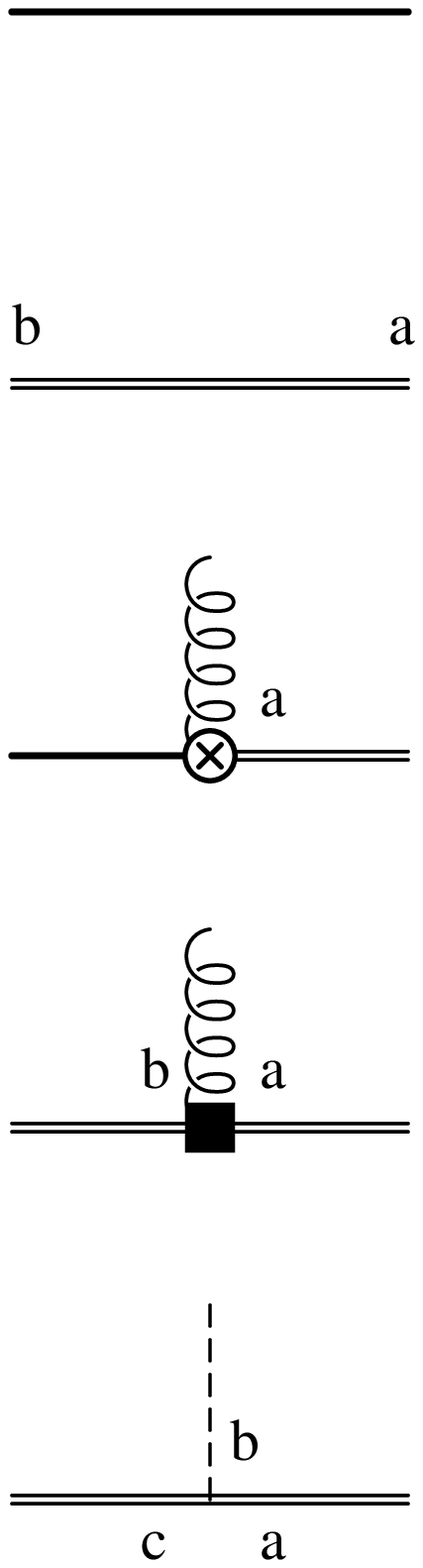}
\put(25,325){$= \theta(T) e^{\displaystyle -i V_s T}$}\put(210,325){singlet propagator}
\put(25,248){$= \theta(T)\Bigg(e^{\displaystyle -i V_o T} e^{\displaystyle - i g \int_{-T/2}^{T/2} \!\!\! dt 
                 \, A_0^{\rm adj}}\Bigg)_{ab}$} \put(210,248){octet propagator}
\put(25,171){$= i g V_A \displaystyle\sqrt{T_F\over N_c} {\bf r} \cdot {\bf E}^a$}\put(210,171){singlet--octet vertex}
\put(25,94){$= i g \displaystyle{V_B\over 2} d^{abc} {\bf r} \cdot {\bf E}^c$}\put(210,94){octet--octet vertex}
\put(25,17){$=  g f^{abc}$}\put(210,17){Coulomb octet--octet vertex}
\caption{ \it Propagators and vertices of the pNRQCD Lagrangian (\ref{pnrqcd0}). 
In perturbative calculations the octet propagator is understood without the
gluonic string, using instead the Coulomb octet--octet vertex. Also the singlet--octet and the octet--octet vertices produce three
diagrams each in perturbative calculations with the Coulomb gauge: one
with
a longitudinal gluon line, one with a transverse gluon line, and one with
both a longitudinal and a transverse gluon line.}
\label{pnrqcdfig}
\end{figure}

In order to interpret the matching potentials  $V_s$ and $V_o$ properly, let us consider, for 
instance, the leading order (in the multipole expansion) equation of motion of the singlet field
$$
i\partial_0 {\rm S} = \left({{\bf p}^2\over m} + V_s(r)\right){\rm S}\,,
$$
where we have included also the kinetic energy, since it contributes at the same order in 
$v$  as $V_s$ to the physical observables of the system. This equation is the  non-relativistic Schr\"odinger 
equation. $V_s$ is the singlet static potential if the neglected terms in the multipole expansion 
are of non-potential type, i.e. if the next relevant scale 
of the system is the energy scale $m v^2 \sim \partial_0 \sim {\bf p}^2/m \sim V_s(r)$.  
If this is not the case, for instance if the next relevant scale is $\Lambda_{\rm QCD}$, the 
neglected terms in the multipole expansion still give potential type contributions 
i.e. contributions that do not depend on the energy/state of the system and  
$V_s$ is not the static singlet potential. Therefore, we can identify the matching potentials 
$V_s$ and $V_o$ with the singlet and octet heavy $Q$-$\bar Q$ 
static potential, respectively, only if no other physical scale appears between the soft scale $m v$ 
and the ultrasoft scale $m v^2$, i.e. if $\Lambda_{\rm QCD} \siml mv^2$. See section \ref{secl1} for 
a discussion of this situation. If this is not the case, in order to define a potential,   
scales larger than the ultrasoft one have to be integrated out. This situation will be discussed 
in section \ref{secl2}. 

We define 
\begin{equation}
V_s(r) \equiv  - C_F {\alpha_{V_s}(r) \over r}, 
\qquad 
V_o(r) \equiv  \left({C_A\over 2} -C_F\right) {\alpha_{V_o}(r) \over r}, 
\label{defpot}
\end{equation}
where $C_F$ and $C_A$ are respectively the Casimir of the fundamental and of
the adjoint representation 
of the gauge group. In QCD $C_F = 4/3$ and $C_A = N_c = 3$.  
$V_A$ and $V_B$ are the matching coefficients associated in the Lagrangian (\ref{pnrqcd0}) 
to the leading corrections in the multipole expansion. The coefficients $V_s$, $V_o$,   
$V_A$ and $V_B$ have to be determined by matching pNRQCD with NRQCD at a scale
$\mu$ of the order of or smaller 
than $m v$, and larger than the next relevant scale. Since we have assumed that $\mu$ is 
larger than $\Lambda_{\rm QCD}$, the matching can be done perturbatively. 
In order to have the proper free-field normalization in the colour space we define 
\begin{equation}
{\rm S} \equiv { 1\!\!{\rm l}_c \over \sqrt{N_c}} S \quad \quad {\rm O} \equiv  { T^a \over \sqrt{T_F}}O^a, 
\label{norm}
\end{equation}
where $T_F=1/2$. $S$ and $O$ are $1/2\times 1/2$ tensors in spin space. The Feynman rules for the propagators 
and vertices defined by the Lagrangian (\ref{pnrqcd0}) are shown in Fig. \ref{pnrqcdfig}.  

In principle, terms with higher time derivatives acting on the fields 
$S$ or $O$ could be considered in the pNRQCD Lagrangian. These terms are 
redundant in the sense that one could make them disappear from the Lagrangian and still correctly predict 
all physical observables by re-shuffling the values of the matching coefficients. We have chosen the minimal form of the pNRQCD Lagrangian, where higher time derivatives are absent. In fact, one can always get rid of these terms by systematically using 
local field redefinitions without changing the physical observables 
(spectrum) of the theory. 

\section{Matching at leading order in $1/m$}
\label{secmatch}
In this section we will discuss how to perform the matching between NRQCD and pNRQCD. 
In particular we will explicitly calculate the matching at leading order in the $1/m$ expansion 
and up to $O(r^2)$ in the multipole expansion. 
 
The matching is in general done by comparing 2-fermion Green functions 
(plus external gluons with energy of $O(\Lambda_{mp})$) in NRQCD and pNRQCD, order by 
order in $1/m$ and order by order in the multipole expansion. This is rigorously justified 
with the choice of cut-offs made in section 2.2. If the soft scale is in 
the perturbative region of QCD (i.e. it is larger than $\Lambda_{\rm QCD}$), this can be done 
in addition order by order in the coupling constant, $\als$. In perturbation theory, matching calculations 
are usually done, for simplicity, using Green functions which are not gauge invariant \cite{NRQCD,Manohar,Match}. 
If the soft scale is not in the perturbative region (i.e. it is  comparable to  or smaller than $\Lambda_{\rm QCD}$)
one can still perform the matching, but a nonperturbative evaluation of 
the Green functions is then required. In this case it is important to consider gauge invariant Green functions only. 
We will try to present formulas for both situations. However, since we will be concerned with 
the situation $1/r > \Lambda_{\rm QCD}$, only the formulas suitable for a perturbative matching 
(which typically are not gauge invariant, but easier to handle) will be used in the actual calculations.

When matching NRQCD to pNRQCD it could eventually happen that one does not directly end up with the 
pNRQCD Lagrangian in its minimal form and some terms with higher time derivatives are needed. 
These subtleties do not affect the matching at the order we are working in this paper.

\subsection{Interpolating fields}
The matching can be done once the interpolating fields for $S$ and $O^a$ have been identified in NRQCD. 
The former need to have the same quantum numbers and the same transformation properties as the latter. 
The correspondence is not one-to-one. Given an interpolating field in NRQCD there is an infinite number of 
combinations of singlet and octet wave-functions with ultrasoft fields which have the same quantum numbers 
and therefore have a non vanishing overlap with the NRQCD operator. Fortunately, the operators in pNRQCD 
can be organized according to the counting of the multipole expansion. 
For instance, for the singlet we have  
\begin{equation}
\chi^\dagger({\bf x}_2,t) \phi({\bf x}_2,{\bf x}_1;t) \psi({\bf x}_1,t) =  Z^{1/2}_s(r) S({\bf R},{\bf r},t) 
+ Z^{1/2}_{E,s}(r) r \, {\bf r}\cdot{\bf E}^a({\bf R},t) O^a({\bf R},{\bf r},t) + \dots,  
\label{Sdef}
\end{equation}
and  for the octet 
\begin{eqnarray}
\!\!\!\chi^\dagger({\bf x}_2,t) \phi({\bf x}_2,{\bf R};t) T^a \phi({\bf R},{\bf x}_1;t) \psi({\bf x}_1,t) 
&=& Z^{1/2}_o(r) O^a({\bf R},{\bf r},t) \nonumber\\
&+& Z^{1/2}_{E,o}(r) r \, {\bf r}\cdot{\bf E}^a({\bf R},t) S({\bf R},{\bf r},t) + \dots, 
\label{Odef}
\end{eqnarray}
where 
\begin{equation}
\phi({\bf y},{\bf x};t)\equiv {\rm P} \exp \left\{ ig \displaystyle 
\int_0^1 \!\! ds \, ({\bf y} - {\bf x}) \cdot {\bf A}({\bf x} - s({\bf x} - {\bf y}),t) \right\}.
\label{schwinger}
\end{equation}
As it is clear from Eqs. (\ref{Sdef}) and (\ref{Odef}) these operators guarantee a
leading overlap with the singlet and the octet wave-functions respectively. Higher order corrections are 
suppressed in the multipole expansion and will not be considered in the applications 
of the following sections. In particular, higher order operators in the expansions (\ref{Sdef}) and (\ref{Odef}) 
contribute to the singlet and octet matching at order $O(r^3)$ or higher in the multipole expansion. 
In the following we will not go beyond effects of order $O(r^2)$. We notice that in the expansions 
(\ref{Sdef}) and (\ref{Odef}) the momentum operator ${\bf p}$ does not appear since we are performing 
the matching at the lowest order in the $1/m$ expansion.

The strings attached to the quarks on the NRQCD side contain soft contributions. Their role 
in the calculation will be either to produce corrections which are exponentially suppressed 
or to change the normalization factors $Z$. They will not change the pole of the correlator. 
This is discussed in detail in appendix A. We will show explicitly how this works 
in the perturbative singlet matching of section \ref{subsecnlompvs}. Note that no assumptions are required 
on the large time behaviour of the gauge fields as it is often claimed  in the literature, 
(like $A_{\mu}(T \rightarrow \infty)=0$, which is, in general, not true).

Notice that the matching for the octet (\ref{Odef}) is not carried out
using gauge invariant operators. 
In a perturbative matching this is not problematic, since then, as we will we see, we expect $V_o$, 
which corresponds to the octet propagator pole, to be gauge invariant order by order 
in $\alpha_{\rm s}$. However, one may prefer to work with manifestly gauge invariant quantities. 
This is the case if eventually one wants to take advantage of nonperturbative techniques like 
lattice simulations. A possible solution consists in substituting the $T^a$ 
colour matrix on the left side of Eq. (\ref{Odef}) by a local gluonic operator $H^a({\bf R},t)T^a$ 
with the right transformation properties. The NRQCD operator has then the form   
\begin{equation}
\chi^\dagger({\bf x}_2,t) \phi({\bf x}_2,{\bf R},t)\, H({\bf R},t) \, \phi({\bf R},{\bf x}_1,t) \psi({\bf x}_1,t).
\label{Ogi}
\end{equation}
To be more specific a possible gauge invariant matching condition for the octet reads 
\begin{equation}
\chi^\dagger({\bf x}_2,t) \phi({\bf x}_2,{\bf R},t)\,{\bf B}({\bf R},t)\,\phi({\bf R},{\bf x}_1,t) \psi({\bf x}_1,t) 
= Z^{1/2}_{B,o}(r) {\bf B}^a({\bf R},t)O^a({\bf R},{\bf r},t) + \dots,
\label{Odefgi}
\end{equation}
where the dots indicate overlaps with operators of higher order in the multipole expansion.
Note that, if we would have identified $H$ in Eq. (\ref{Ogi}) with a chromoelectric field, the matching 
condition equivalent to (\ref{Odefgi}) would have produced  an overlap with the singlet operator,  
$\displaystyle {{\bf r}\over r^3} S$, stronger than with the octet one, ${\bf E}^a O^a$.  
From the NRQCD point of view this is due to the fact that in Eq. (\ref{Ogi}) nothing prevents 
the dynamical gluons associated with the inserted operator to be soft just as it happens 
for the gluonic strings attached to the quarks.

\subsection{Matching at order $r^0$ in the Multipole Expansion}
\label{seclm}
In this section we identify the matching operators relevant for the matching of $V_s$, $Z_s$, $V_o$ and $Z_o$
at order $(1/m)^0$ in the mass expansion and at order $r^0$ in the multipole expansion.

\subsubsection{Singlet Matching}
\begin{figure}[htb]
\makebox[5.0cm]{\phantom b}
\epsfxsize=6truecm \epsfbox{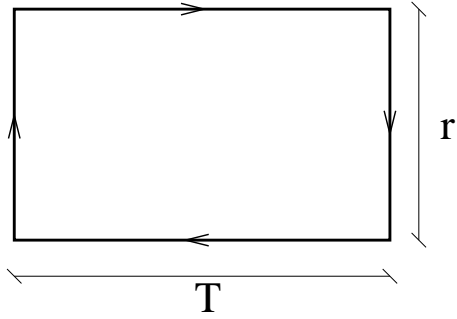}
\caption{ \it A graphical representation of the static Wilson loop. We adopt the convention 
that time propagates from the left to the right. Therefore, the quarks trajectories 
are represented by horizontal lines and the equal-time endpoint Schwinger strings 
by shorter vertical lines.}
\label{wilsonfig}
\end{figure}

In order to get the singlet matching potential $V_s$ and the singlet
normalization factor $Z_s$, 
we choose the following Green function: 
\begin{equation}
I_s \equiv \langle 0 \vert  \chi^\dagger(x_2) \phi(x_2,x_1) \psi(x_1) 
\psi^\dagger(y_1)\phi(y_1,y_2) \chi(y_2) \vert 0 \rangle. 
\label{vsmatch}
\end{equation}
In NRQCD we obtain 
\begin{equation}
I_s = \delta^3({\bf x}_1 - {\bf y}_1) \delta^3({\bf x}_2 - {\bf y}_2) \langle W_\Box \rangle , 
\label{vsnrqcd}
\end{equation}
where $W_\Box$ is the rectangular Wilson loop of Fig. \ref{wilsonfig} 
with edges $x_1 = (T/2,{\bf r}/2)$, $x_2 = (T/2,-{\bf r}/2)$, 
$y_1 = (-T/2,{\bf r}/2)$ and  $y_2 = (-T/2,-{\bf r}/2)$. 
The brackets  $\langle ~~ \rangle$ stand for  the average over the gauge
fields and light quarks.

\begin{figure}[htb]
\makebox[0.2cm]{\phantom b}
\epsfxsize=16truecm \epsfbox{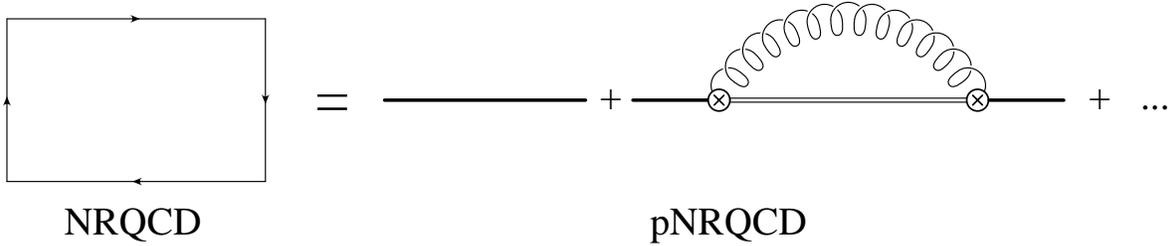}
\vspace{0.2cm}
\caption{ \it The matching of $V_s$ and $Z_s$ at next-to-leading order in the multipole 
expansion. On the left-hand side is the Wilson loop in NRQCD, on the right-hand  
side are the pNRQCD fields. The wavy line represents the ultrasoft gluon propagator.}
\label{figmats}
\end{figure}

Equation (\ref{Sdef}) states that the leading overlap of the Green function (\ref{vsnrqcd}) 
in pNRQCD is with the singlet propagator, see also Fig. \ref{figmats}. Indeed, in pNRQCD 
we obtain at the zeroth order in the multipole expansion:
\begin{equation}
I_s = Z_s(r) \delta^3({\bf x}_1 - {\bf y}_1) \delta^3({\bf x}_2 - {\bf y}_2) e^{-iTV_s(r)}.
\label{vspnrqcd}
\end{equation}

$\langle W_\Box \rangle$ is in general a complicated function of $T$, but we just want to single out the 
soft scale. This means that we only need to know it in the limit $T \rightarrow \infty$, i.e. 
as a $1/T$ expansion, in the NRQCD calculation. We define
\begin{equation}
{i\over T}\ln  \langle W_\Box \rangle = u_0(r) + i {u_1(r)\over T} + O\left( {1\over T^2}\right) 
\quad \hbox{for} \quad T\to\infty \,   .
\label{sexp}
\end{equation}
Therefore, comparing Eq. (\ref{vsnrqcd}) with  Eq. (\ref{vspnrqcd}), we get 
\begin{eqnarray}
V_s(r) &\equiv& -C_F {\alpha_{V_s}(r) \over r} = u_0(r), \label{vs1}\\
\ln Z_s(r) &=& u_1(r). \label{zs1}
\end{eqnarray}

Note that the matching above does not rely on any perturbative expansion in 
$\alpha_{\rm s}$. However, since we are concerned with the situation $r < 1/\Lambda_{\rm QCD}$ 
the matching can be done in addition perturbatively, i.e. the quantities on the right-hand side 
of Eqs. (\ref{vs1}) and (\ref{zs1}) can be evaluated expanding order by order in $\alpha_{\rm s}$. 
At leading order in $\alpha_{\rm s}$ we get
\begin{eqnarray}
V_s(r) &=& -C_F {\alpha_{\rm s} \over r} \quad \hbox{or} \quad \alpha_{V_s} =  \alpha_{\rm s}, \label{vs1leading}\\
Z_s(r) &=& N_c. \label{zs1leading}
\end{eqnarray}

\subsubsection{Octet Matching} 
\label{secvoma}
In order to get the octet matching potential $V_o$ and the octet normalization
factor $Z_o$, we choose the following Green function: 
\begin{eqnarray}
I_o^{ab} &\equiv& 
\langle 0 \vert  \chi^\dagger(x_2) \phi\left({\bf x}_2,{{\bf x}_1+{\bf x}_2\over 2};{T\over 2}\right) 
T^a \phi\left({{\bf x}_1+{\bf x}_2\over 2},{\bf x}_1;{T\over 2}\right) \psi(x_1) \nonumber\\
& & \qquad\qquad \times \psi^\dagger(y_1)\phi\left({\bf y}_1,{{\bf y}_1+{\bf y}_2\over 2};-{T\over 2}\right) 
T^b \phi\left({{\bf y}_1+{\bf y}_2\over 2},{\bf y}_2;-{T\over 2}\right) \chi(y_2) \vert 0 \rangle. 
\label{vomatch}
\end{eqnarray}
In NRQCD we have 
\begin{equation}
I_o^{ab} = \delta^3({\bf x}_1 - {\bf y}_1) \delta^3({\bf x}_2 - {\bf y}_2) \langle T^a W_\Box T^b \rangle, 
\label{vonrqcd}
\end{equation}
where the colour matrices are intended as path ordered insertions on the static 
Wilson loop in the points (${\bf R},T/2$) and (${\bf R},-T/2$). 

Eq. (\ref{Odef}) states that the leading overlap of the Green function (\ref{vonrqcd}) 
in pNRQCD is  with the octet propagator. Indeed, in pNRQCD we obtain 
at the zeroth order in the multipole expansion:
\begin{equation}
I_o^{ab} = Z_o(r) \delta^3({\bf x}_1 - {\bf y}_1) \delta^3({\bf x}_2 - {\bf y}_2) 
e^{-iTV_o(r)} \langle \phi(T/2,-T/2)_{ab}^{\rm adj} \rangle,  
\label{vopnrqcd}
\end{equation}
where the Schwinger line  
$$
\phi(T/2,-T/2) \equiv \phi(T/2,{\bf R},-T/2,{\bf R}) 
= {\rm P} \exp \left\{ - ig \displaystyle \int_{-T/2}^{T/2} \!\! dt \, A_0({\bf R},t) \right\}
$$
is evaluated in the adjoint representation.

As in the singlet case, we only need to know the $1/T$ expansion of $\langle T^a W_\Box T^b \rangle$.  
We define 
\begin{equation}
{i\over T}\ln  {\langle T^a W_\Box T^b \rangle \over \langle \phi(T/2,-T/2)_{ab}^{\rm adj}\rangle}
= v_0(r) + i {v_1(r)\over T} + O\left( {1\over T^2}\right) 
\quad \hbox{for} \quad T\to\infty \,   .
\label{soexp}
\end{equation}
Therefore, comparing Eq. (\ref{vonrqcd}) with  Eq. (\ref{vopnrqcd}), we get 
\begin{eqnarray}
V_o(r) &\equiv& \left({C_A\over 2} -C_F\right) {\alpha_{V_o}(r) \over r} = v_0(r), \label{vo1}\\
\ln Z_o(r) &=& v_1(r). \label{zo1}
\end{eqnarray}
Again, the formulas above do not rely on any expansion in $\alpha_{\rm s}$.
If in addition the matching can be done perturbatively ($r < 1/\Lambda_{\rm QCD}$),  
at leading order in $\alpha_{\rm s}$ we get
\begin{eqnarray}
V_o(r) &=& \left({C_A\over 2} - C_F\right) {\alpha_{\rm s} \over r} \quad \hbox{or} \quad 
\alpha_{V_o} =  \alpha_{\rm s}, \label{vo1leading}\\
Z_o(r) &=& T_F. \label{zo1leading}
\end{eqnarray}
 
\begin{figure}[htb]
\makebox[0.2cm]{\phantom b}
\epsfxsize=16truecm \epsfbox{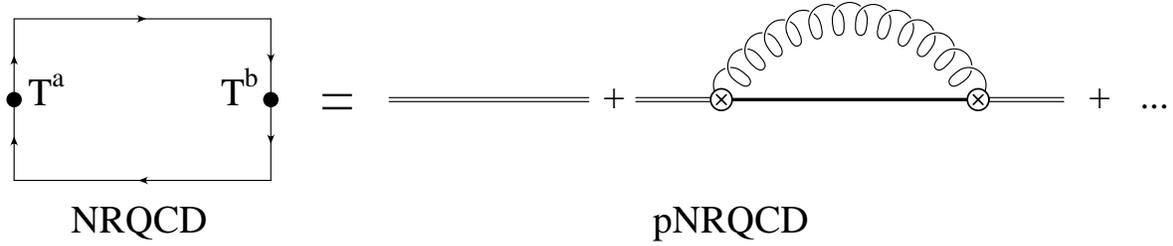}
\vspace{0.2cm}
\caption{ \it The matching of $V_o$ and $Z_o$ at next-to-leading order in the multipole 
expansion. On the left-hand side is the Wilson loop in NRQCD with colour matrices insertions, 
on the right-hand side are the pNRQCD fields.}
\label{figmato}
\end{figure}

The matching discussed above is gauge dependent. Still, the octet matching potential defined above is expected to be a gauge independent quantity at any finite order in 
perturbation theory, since it corresponds to the pole of the propagator, i.e. 
to an eigenvalue of the spectrum (just in the same way as the pole mass 
of the quark propagator is a gauge independent quantity at any finite order 
in perturbation theory \cite{Breckenridge}). At this respect it is easy to see on 
dimensional grounds (by Fourier transforming $T$ to energy space) 
that the string $\langle \phi(T/2,-T/2)_{ab}^{\rm adj}\rangle$ does 
not give contribution to the potential at any finite order in perturbation theory, 
although it does to $Z_o$, which turns out to be a gauge dependent object. 

The octet matching can also be obtained in a gauge 
invariant way by choosing, for instance, the following gauge invariant Green function
\begin{eqnarray}
I_o^{g.i.} &\equiv& \langle 0 \vert  \chi^\dagger(x_2) 
\phi\left({\bf x}_2,{{\bf x}_1+{\bf x}_2\over 2};{T\over 2}\right) 
{\bf B}\left({{\bf x}_1+{\bf x}_2\over 2};{T\over 2}\right) 
\phi\left({{\bf x}_1+{\bf x}_2\over 2},{\bf x}_1;{T\over 2}\right) \psi(x_1) \nonumber\\
& & \!\!\!\!\!\!\!\!\!\!\!\!\!\!\!
\times \psi^\dagger(y_1)\phi\left({\bf y}_1,{{\bf y}_1+{\bf y}_2\over 2};-{T\over 2}\right) 
{\bf B}\left({{\bf y}_1+{\bf y}_2\over 2};-{T\over 2}\right) 
\phi\left({{\bf y}_1+{\bf y}_2\over 2},{\bf y}_2;-{T\over 2}\right) \chi(y_2) \vert 0 \rangle. 
\label{vomatchgi}
\end{eqnarray}
Eq. (\ref{Odefgi}) states that the leading overlap of the Green function (\ref{vomatchgi}) 
in pNRQCD is  with the octet propagator. In a way similar to  the gauge dependent 
matching, we define for $T\to\infty$ 
\begin{equation}
{i\over T}\ln  {\langle {\bf B}({\bf R},T/2) W_\Box {\bf B}({\bf R},-T/2) \rangle
\over \langle {\bf B}^a({\bf R},T/2) \phi(T/2,-T/2)_{ab}^{\rm adj} {\bf B}^b({\bf R},-T/2)\rangle}
= v_0^{g.i.}(r) + i {v_1^{g.i.}(r)\over T} + O\left( {1\over T^2}\right) 
\label{soexpgi}
\end{equation}
and we get 
\begin{eqnarray}
V_o(r) &=& v_0^{g.i}(r), \label{vo1gi}\\
\ln Z_{B,o}(r) &=& v_1^{g.i.}(r). \label{zo1gi}
\end{eqnarray}
We note that $V_o$ has to be independent of the operator used as far as it has a leading 
projection on the octet state, while in general the octet normalization factor
is not.

\subsection{Matching at order $r^1$ in the multipole expansion}
At $O(r)$ there are no additional contributions to the singlet and octet matching potentials 
and to the normalization factors. At this order one finds the leading contributions 
to the mixed potentials $V_A$ and $V_B$. For the purposes of this paper they are only needed 
to be known perturbatively at leading order in $\alpha_{\rm s}$, 
$$
V_A(r) = 1,  \qquad\qquad V_B(r) = 1.
$$

\subsection{Matching at order $r^2$ in the multipole expansion}
At this order we find the next-to-leading contributions to the singlet and octet potentials 
and to the singlet normalization factor.

\subsubsection{Singlet Matching}
\label{subsecnlompvs}
The next-to-leading order correction in the multipole expansion to 
the singlet matching (\ref{vspnrqcd}) is given by (see Fig. \ref{figmats}) 
\begin{eqnarray}
\label{vspnrqcdus}
& &\!\!\!\!\!\!\! I_s = Z_s(r) \delta^3({\bf x}_1 - {\bf y}_1) \delta^3({\bf x}_2 - {\bf y}_2) e^{-iTV_s(r)} 
\\
\nonumber
& &\!\!\!\!\!\!\! \times \left(1 -{ g^2 \over N_c} T_F V_A^2 (r)
\int_{-T/2}^{T/2} \! dt \int_{-T/2}^{t} \! dt^\prime \, e^{-i(t-t^\prime)(V_o-V_s)} 
\langle {\bf r}\cdot {\bf E}^a(t) \phi(t,t^\prime)^{\rm adj}_{ab}{\bf r}\cdot {\bf E}^b(t^\prime)\rangle\right)\,,
\end{eqnarray}
where fields with only temporal arguments are evaluated in the centre-of-mass coordinate 
(we will use this notation in the rest of the work). Comparing Eqs. (\ref{vsnrqcd}) and (\ref{vspnrqcdus}) (for large $T$), one
gets the singlet normalization factor $Z_s$ and the singlet matching potential $V_s$  at the next-to-leading 
order in the multipole expansion. Eq. (\ref{vspnrqcdus}) does not rely on any perturbative 
expansion in $\alpha_{\rm s}$. However, since we are considering the case where the matching 
scale $\mu$ is much larger than $ \Lambda_{\rm QCD}$
the equality (\ref{vspnrqcdus}) can be evaluated order by order in $\alpha_{\rm s}$.
Note that this happens  even when the correlator on the right-hand side 
of Eq. (\ref{vspnrqcdus}) in the physical system is dominated by nonperturbative effects.
We shall illustrate the use of Eq. (\ref{vspnrqcdus}) by calculating the leading 
log contribution to the static potential. This will be done in two different ways.

\medskip

The first way consists in  a suitable modification of perturbation theory which is suggested by the 
right-hand side of (\ref{vspnrqcdus}). $\alpha_{\rm s}$ appears in the second term explicitly 
in $V_o$ and $V_s$ (also in $V_A$ at higher orders) and implicitly in the correlator of gluonic operators. 
We shall expand the implicit dependence, i.e. the gluonic correlator, and keep the remaining dependence, 
i.e. $V_o-V_s$, unexpanded. At tree level (in dimensional regularization) we have 
\begin{equation}
T_F \langle E^a_i(t) \phi(t,t^\prime)^{\rm adj}_{ab}E^b_i(t^\prime)\rangle 
= {d-2 \over 2} C_F C_A \mu^{4-d} \int {d^{d-1}k\over {(2\pi)^{d-1}}} k e^{-ik|t-t^\prime|}. 
\label{mgeq0}
\end{equation}
Using the identity (in Euclidean space)
$$
\int{d^d k\over(2\pi)^d}{k^n\over (a+k)^{\nu}} = (-1)^\nu {2\pi^{d\over 2}\over\Gamma(d/2)} 
{1\over(2\pi)^d}{\pi\over\sin(\pi(d+n-1))}{\Gamma(d+n)\over\Gamma(d+n-\nu+1)\Gamma(\nu)}a^{d+n-\nu}\,    ,
$$
we get in the $T\to\infty$ limit ($d = 4+2\epsilon$)
\begin{eqnarray}
& & I_s = Z_s(r) \delta^3({\bf x}_1 - {\bf y}_1) \delta^3({\bf x}_2 - {\bf y}_2) e^{-iTV_s(r)}
\nonumber\\
& & \times \Biggl(1 +i T C_F{ \als \over \pi} { r^2 \over 3} 
(V_o-V_s)^3 \left\{ {1 \over \epsilon} + \gamma_E + \ln{(V_o-V_s)^2 \over 4\pi\mu^2} + {\rm constant} \right\} 
\nonumber\\
& & + C_F{ \als \over \pi} r^2  (V_o-V_s)^2 
\left\{ {1 \over \epsilon} + \gamma_E + \ln{(V_o-V_s)^2 \over 4\pi\mu^2} + {\rm constant} \right\}
+ O\left( {1\over T} \right) \Biggr). 
\label{vspnrqcdusp}
\end{eqnarray}
where $V_A$ has been substituted by its tree level value and
 $V_o-V_s$ is also meant to be substituted by it. 
Note that the UV divergences can be re-absorbed by a renormalization of $V_s$ and $Z_s$. 
\begin{figure}[htb]
\makebox[0.5cm]{\phantom b}
\epsfxsize=15truecm \epsfbox{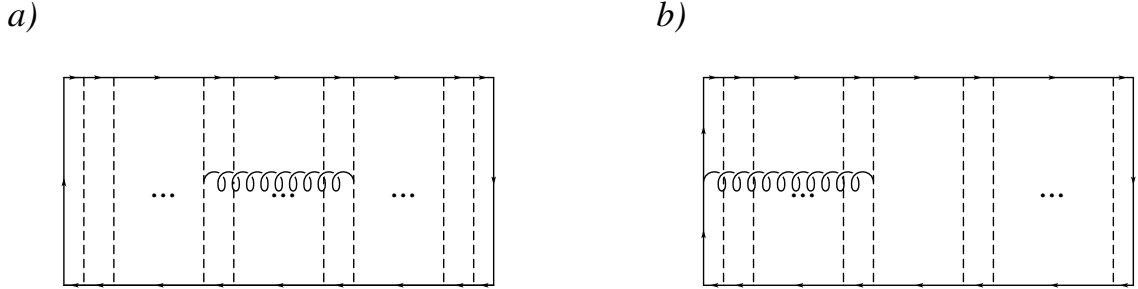}
\vspace{0.2cm}
\caption{ \it Graphs contributing to $\langle W_\Box \rangle$ giving rise 
to $T \alpha_{\rm s}^4\ln{r\,(V_o-V_s)}$ and to $\alpha_{\rm s}^3\ln{r\,(V_o-V_s)}$ terms. 
Dashed lines represent Coulomb exchanges.}
\label{figapp}
\end{figure}
Eq. (\ref{vspnrqcdusp}) also contains the leading logarithm dependence on $\mu$ of $V_s$ and $Z_s$. 
Here we want to check explicitly the cancellation of the 
$\ln (V_o-V_s)$ dependence, which plays the role of an IR cut-off,
with the right-hand side of Eq. (\ref{vsnrqcd}) by calculating at the same order  
the relevant graphs of $\langle W_\Box \rangle$. These are shown in Fig. \ref{figapp}. 
The resummation of ladders (longitudinal gluon propagators in the Coulomb gauge) 
amounts to keep $V_o-V_s$ unexpanded. The graphs {\it a)} were first discussed 
in \cite{Appelquist} and calculated in \cite{BPSV}. 
The graphs {\it b)} (and the symmetric ones) are calculated here for the first time. 
They involve the end-point strings and contribute to the normalization factor only. 
We get for $T\to\infty$ 
\begin{eqnarray*}
\!&{\it a)}& \!\!\delta {i\over T} \ln \langle W_\Box \rangle = 
-{C_FC_A^3\over 24} {\alpha_{\rm s}\over r}{\alpha_{\rm s}^3\over \pi}\ln((V_o-V_s)^2 r^2)
+{i\over T}{C_FC_A^2\over 12} {\alpha_{\rm s}^3\over \pi}\ln ((V_o-V_s)^2 r^2)
+O\left({1\over T^2}\right),\\
\!&{\it b)}& \!\!\delta {i\over T} \ln \langle W_\Box \rangle = 
{i\over T}{C_FC_A^2\over 6} {\alpha_{\rm s}^3\over \pi}\ln ((V_o-V_s)^2 r^2)
+O\left({1\over T^2}\right).
\end{eqnarray*} 
Comparing Eq. (\ref{vsnrqcd}) with Eq. (\ref{vspnrqcdus}) we get ($V_o -V_s = C_A\alpha_{\rm s}/2 r$ at tree level)
\begin{eqnarray}
V_s(r,\mu) &=& -C_F {\alpha_{V_s}(r) \over r} = (u_0(r))_{\rm two-loops} 
-{C_FC_A^3\over 12} {\alpha_{\rm s}\over r}{\alpha_{\rm s}^3\over \pi}\ln{r\mu},
\label{vsus}\\
\ln Z_s(r,\mu) &=& (u_1(r))_{\rm two-loops} + {C_FC_A^2\over 2} {\alpha_{\rm s}^3\over \pi}\ln{r\mu}.
\label{zsus}
\end{eqnarray}
$(u_0(r))_{\rm two-loops}$ is known \cite{twoloop}. See section \ref{secl1} for a discussion. 
Note that the new contributions given in Eq. (\ref{vsus}) and (\ref{zsus}) would be zero in QED. 
The fact that $\alpha_{V_s}$ depends on the IR behaviour of the theory
is, therefore, a distinct feature of QCD. Moreover, we stress that in order to 
match the normalization factor (\ref{zsus}) it is necessary 
to take into account contributions coming from the end-point strings, which   
can be considered irrelevant only at order $T^0$, i.e. for the matching of the potential. 

\medskip 

The second way of carrying out the calculation is by straightforward perturbation theory 
in $\alpha_{\rm s}$, and is standard when doing matching calculations (see \cite{Manohar,pNRQCD,Beneke2,Match,pos}).
If we expand in $\alpha_{\rm s}$ and in the ultrasoft scale ($V_o-V_s$),
both expressions (\ref{vsnrqcd}) and (\ref{vspnrqcdus}) are IR-divergent (Eq. (\ref{vspnrqcdus}) 
is also UV-divergent). By regulating both expressions in dimensional regularization, 
the calculation in pNRQCD gives zero (there is no scale), while the calculation of the Wilson loop shows 
up an explicit dependence on the infrared regulator $\mu$ via a typical $\ln \mu r$ term.
In this situation the matching reduces to enforce the following equality order by order in $\als$ (for large $T$)
\begin{equation}
Z_s(r)e^{-iV_sT}=\langle W_\Box \rangle \,,
\label{matchingper}
\end{equation}
where the exponential has to be perturbatively expanded in $\alpha_{\rm s}$.
In the right-hand side of Eq. (\ref{matchingper}) the first IR divergence 
appears at order $O(\alpha_{\rm s}^3)$ at the two-loop level and 
it only affects the normalization factor. It comes from the diagram {\it a)} of Fig. \ref{figapp} without 
any Coulomb exchange except the ones connected to the transverse gluon and from the diagram {\it b)} 
of Fig. \ref{figapp} with only one extra Coulomb exchange besides the ones attached to the transverse gluon. 
The leading log correction to the normalization factor reads
$$
Z_s(r)\equiv C_A\left(1+\delta Z_s(r)\right)\,, 
\quad \delta Z_s(r)={C_FC_A^2 \over 2}{\alpha_{\rm s}^3 \over \pi}\ln{r\mu}.
$$
We can see that the scale dependence coincides with the calculation previously 
performed in the effective theory (see Eq. (\ref{zsus})).
At order $O(\alpha_{\rm s}^4)$ (three-loops) we have the following equality 
$$
C_A \left(iT\delta Z_s(r){C_F\alpha_{\rm s} \over r}-iT\delta V_s \right)= \delta \langle W_\Box \rangle,
$$
where the computation on the NRQCD side corresponds to the diagrams {\it a)} of Fig. \ref{figapp} 
with only one extra Coulomb exchange besides the ones attached to the transverse gluon and 
$\delta V_s$ stands for the leading logarithm correction to $V_s$. We obtain for $\delta \langle W_\Box \rangle$ 
(up to terms which do not contribute to the potential)
$$
\delta \langle W_\Box \rangle = i { \alpha_{\rm s}^4 \over \pi} T C_A { C_F \over r} {C_A^2 \over 6} 
\left( C_F +{C_A \over 2} \right) \ln{r\mu}.
$$
We can now obtain the leading logarithm correction to $V_s$, or equivalently 
to $\alpha_{V_s}$. As expected we get the same result as that one reported in Eq. (\ref{vsus}).

\subsubsection{Octet Matching} 
\label{subsecnlompvs2}
\begin{figure}[htb]
\makebox[5cm]{\phantom b}
\epsfxsize=6.5truecm \epsfbox{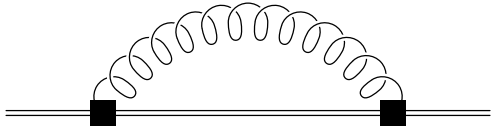}
\vspace{0.2cm}
\caption{ \it Octet self-energy graph.}
\label{figzoo}
\end{figure}

The next-to-leading correction to Eq. (\ref{vopnrqcd}) in the multipole expansion 
is given by (see Fig. \ref{figmato})
\begin{eqnarray}
I_o^{ab} &=& Z_o(r) \delta^3({\bf x}_1 - {\bf y}_1) \delta^3({\bf x}_2 - {\bf y}_2) 
e^{-iTV_o(r)} \Bigg( \langle \phi(T/2,-T/2)_{ab}^{\rm adj} \rangle   
\label{vopnrqcdus}\\ 
& & \!\!\!\!\!\!\!\!\!\!\!\!\!\!\!\!\!\!\! 
- { g^2 \over N_c} T_F V_A^2 (r) \int_{-T/2}^{T/2} \!\! dt 
\int_{-T/2}^{t} \! dt^\prime e^{-i(t-t^\prime)(V_s-V_o)} 
\langle \phi(T/2,t)^{\rm adj}_{aa^\prime} {\bf r}\cdot {\bf E}^{a^\prime}(t) 
{\bf r}\cdot {\bf E}^{b^\prime}(t^\prime) \phi(t^\prime,-T/2)^{\rm adj}_{b^\prime b}\rangle\Bigg). 
\nonumber
\end{eqnarray}
We have omitted above a term proportional to $V_B^2$ (see Fig. \ref{figzoo}) and terms which contain 
operators like ${\rm Tr} ({\bf r}^i{\bf r}^j[{\bf D}^i,{\bf E}^j]{\rm O} {\rm O}^{\dagger})$, 
which should also be included in the pNRQCD Lagrangian at order $O(r^2)$, because they 
neither contribute to the octet matching potential nor to the normalization in perturbation theory. 
Hence formula (\ref{vopnrqcdus}) is less general than the analogous formula (\ref{vspnrqcdus}) 
for the singlet matching potential and cannot be used beyond perturbation theory.

We use Eq. (\ref{vopnrqcdus}) in order to calculate the leading log dependence of the octet matching potential. 
From what we have learnt in the previous section, it is sufficient to calculate the leading log dependence 
of the gluon correlator in the right hand side of Eq. (\ref{vopnrqcdus}).
Up to a color factor the contribution of Eq. (\ref{vopnrqcdus}) is equal, at leading order in 
$\alpha_{\rm s}$, to the one calculated in the singlet case (see Eq. (\ref{vspnrqcdus})). Proceeding along 
the same line as for the singlet case we obtain 
\begin{equation}
V_o(r,\mu) = \left({C_A\over 2} -C_F\right) {\alpha_{V_o}(r) \over r} = (v_0(r))_{\rm two-loops} 
+\left({C_A\over 2} -C_F\right) {C_A^3\over 12} {\alpha_{\rm s}\over r}{\alpha_{\rm s}^3\over \pi}\ln{r\mu}. 
\label{vous}
\end{equation}
Up to now $(v_0(r))_{\rm two-loops}$ is unknown. Nevertheless Eq. (\ref{vous}) shows that 
the ultrasoft corrections to $V_o$ are of relative order $O(\alpha_{\rm
  s}^3)$. This makes clear that the widespread believe that $V_o$ suffers from IR divergences at relative order $O(\alpha_{\rm s}^2)$ \cite{wrongoctetpot,Appelquist} 
(likely due to some mixing up with IR divergences of the normalization factor) is not correct. 
We do not present the calculation of the normalization factor $Z_o$, since, as already 
discussed in section \ref{secvoma}, it is a gauge dependent object.

\subsection{One-loop Running}
\label{secrunning}
As a byproduct of the next-to-leading order calculations done above, we have got 
the leading running in the matching scale $\mu$ of the matching potentials 
$V_s$ and $V_o$. For completeness we give here also the running of $V_A$ and $V_B$.

\begin{figure}[htb]
\makebox[0.5cm]{\phantom b}
\epsfxsize=15truecm \epsfbox{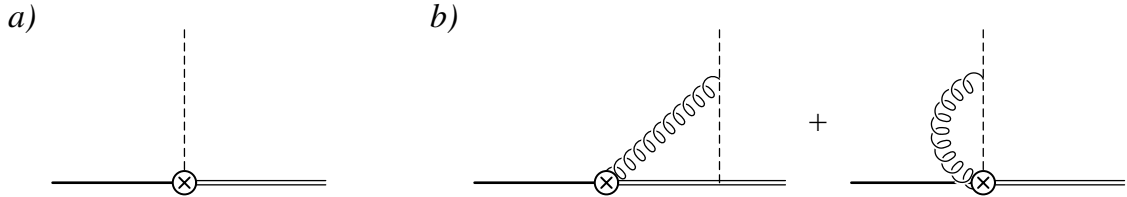}
\vspace{0.2cm}
\caption{ \it Singlet-octet vertex at tree level {\it a)} and at one-loop {\it b)}.}
\label{figva}
\end{figure}

The coefficient $V_A$ controls the running of the singlet-octet vertex 
shown in diagram {\it a)} of Fig. \ref{figva}. Diagram {\it b)} shows the one-loop correction 
to the vertex. The divergent part of it gives the one-loop $\mu$ dependence of $V_A$:
\begin{equation}
V_A(r,\mu) = 1 + {8\over 3} C_A {\alpha_{\rm s}\over \pi} \ln{r\mu}. 
\label{varun}
\end{equation}

\begin{figure}[htb]
\makebox[0.5cm]{\phantom b}
\epsfxsize=15truecm \epsfbox{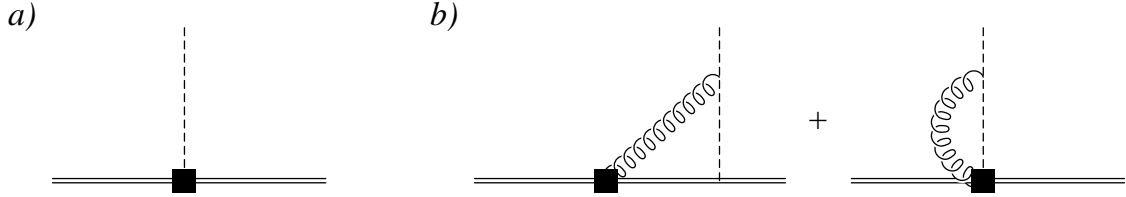}
\vspace{0.2cm}
\caption{ \it Octet-octet vertex at tree level {\it a)} and at one-loop {\it b)} (symmetric graphs are understood).}
\label{figvb}
\end{figure}

Analogously the coefficient $V_B$ controls the running of the octet-octet vertex 
shown in diagram {\it a)} of Fig. \ref{figvb}. Diagram {\it b)} shows the one-loop correction to the vertex 
and the divergent part of it gives the one-loop $\mu$ dependence of $V_B$:
\begin{equation}
V_B(r,\mu) = 1 + {8\over 3} C_A {\alpha_{\rm s}\over \pi} \ln{r\mu}, 
\label{vbrun}
\end{equation}
which turns out to be equal to $V_A$.  In Tab. \ref{tab1} we summarize the 
running of the matching potentials in the pNRQCD Lagrangian at order $O(r)$ 
and of the singlet normalization factor at leading order in perturbation theory.

\begin{table}[ht]
\makebox[4.8cm]{\phantom b}
\begin{tabular}{|c|c|}
\hline
&\\
singlet& $\displaystyle\mu{d \alpha_{V_s}\over d\mu} = {C_A^3\over 12}{\alpha_{\rm s}^4\over\pi}$\\
singlet& $\displaystyle \mu{d Z_s\over d\mu} = {C_FC_A^3\over 2}{\alpha_{\rm s}^3\over\pi} $\\
octet& $\displaystyle \mu{d \alpha_{V_o}\over d\mu} = {C_A^3\over 12}{\alpha_{\rm s}^4\over\pi} $\\
singlet-octet &$\displaystyle \mu{d V_A \over d\mu} = {8\over 3} C_A {\alpha_{\rm s}\over \pi}$\\
octet-octet &$\displaystyle \mu{d V_B \over d\mu} = {8\over 3} C_A {\alpha_{\rm s}\over \pi}$\\
&\\
\hline
\end{tabular}
\caption{ \it One-loop running of the matching potentials of the pNRQCD
  Lagrangian (\ref{pnrqcd0}) and of the singlet normalization factor.}
\label{tab1}
\end{table}

\section{pNRQCD: $\Lambda_{\rm QCD} \siml mv^2$}
\label{secl1}
In the situation $\Lambda_{\rm QCD} \siml mv^2$ there is no relevant physical scale between $m v$ 
and $m v^2$ and the pNRQCD Lagrangian of Eq. (\ref{pnrqcd0}) only describes
ultrasoft degrees of freedom. 
All nonperturbative contributions are encoded into non-potential terms, while 
the static potentials (singlet and octet) coincide with the matching potentials 
$V_s$ and $V_o$ calculated above, i.e. they are purely perturbative. 

In particular, for what concerns the singlet static potential, the result 
(\ref{vsus}) improves the knowledge of it  up to the three-loop leading logarithm order, 
$\alpha_{\rm s}^3 \ln{r \mu}$. Substituting $(u_0(r))_{\rm two-loops}$ by its actual value, we get 
\begin{eqnarray}
&&{\alpha}_{V_s}(r, \mu)=\alpha_{\rm s}(r)
\left\{1+\left(a_1+ 2 {\gamma_E \beta_0}\right) {\alpha_{\rm s}(r) \over 4\pi}\right.
\nonumber\\
&&+\left[\gamma_E\left(4 a_1\beta_0+ 2{\beta_1}\right)+\left( {\pi^2 \over 3}+4 \gamma_E^2\right) 
{\beta_0^2}+a_2\right] {\alpha_{\rm s}^2(r) \over 16\,\pi^2}
\left. + {C_A^3 \over 12}{\alpha_{\rm s}^3(r) \over \pi} \ln{ r \mu}\right\},
\label{newpot}                  
\end{eqnarray}
where $\alpha_{\rm s}$ is in the $\overline{\rm MS}$ scheme, $\beta_n$ are the coefficients of the beta function, 
$a_1$ was calculated in Ref. \cite{1loop} and $a_2$ in Ref. \cite{twoloop} (see \cite{twoloop} for notation). 
One of the most relevant consequences of Eq. (\ref{newpot}) is that $\alpha_{V_s}$ is not a short-distance 
quantity as $\alpha_{\rm s}$ (in the ${\rm \overline{MS}}$ scheme), since it depends on $\mu$. 
Actually, it can be better understood as a matching coefficient. Similar 
conclusions would follow for $\al_{V_o}$, although it is not known with the same accuracy.

The pNRQCD Lagrangian can be systematically written in a $1/m$ expansion. In
this work we are just studying the  $1/m^0$ term. Nevertheless, for the sake
of completeness, we would like to show 
how the Lagrangian looks like when $1/m$ terms are included: 
\begin{eqnarray}                         
& & {\cal L}_{\rm pNRQCD} = {\rm Tr} \,\Biggl\{ {\rm S}^\dagger \left( i\partial_0 
- {{\bf P}^2\over 4m}- {{\bf p}^2\over m} +{{\bf p}^4\over 4m^3}
- V_s(r) - {V_s^{(1)} \over m}- {V_s^{(2)} \over m^2}+ \dots  \right) {\rm S}
\nonumber
\\
&&
\nonumber
\qquad 
+ {\rm O}^\dagger \left( iD_0 - {{\bf P}^2\over 4m} - {{\bf p}^2\over m}+{{\bf p}^4\over 4m^3} 
- V_o(r) - {V_o^{(1)} \over m}- {V_o^{(2)} \over m^2}+\dots  \right) {\rm O} \Biggr\}
\nonumber\\
\nonumber
& &\qquad + g V_A ( r) {\rm Tr} \left\{  {\rm O}^\dagger {\bf r} \cdot {\bf E} \,{\rm S}
+ {\rm S}^\dagger {\bf r} \cdot {\bf E} \,{\rm O} \right\} 
+ g {V_B (r) \over 2} {\rm Tr} \left\{  {\rm O}^\dagger {\bf r} \cdot {\bf E} \, {\rm O} 
+ {\rm O}^\dagger {\rm O} {\bf r} \cdot {\bf E}  \right\}  
\\
& &\qquad- {1\over 4} F_{\mu \nu}^{a} F^{\mu \nu \, a}\,,
\label{pnrqcdph}
\end{eqnarray}
where ${\bf P}$ is the momentum associated to the centre-of-mass coordinate.
In Eq. (\ref{pnrqcdph}) the  $1/m$ corrections to $V_A$, $V_B$ and to pure gluonic operators as well 
as the higher order terms in the multipole expansion are not displayed.
Let us note that the kinetic energy is unavoidable 
when computing physical observables, since $i\partial_0\sim {\bf p}^2/m \sim V_s(r)\sim mv^2$. 
The different $V^{(n)}_{s,o}$ can depend on ${\bf r}$, ${\bf p}$ and the spin. 
They are purely perturbative quantities, just in the same way as $V_{s,o}$ is and 
have been computed with some degree of accuracy in the literature.

With (\ref{pnrqcdph}) we have finally obtained the relevant Lagrangian for
ultrasoft degrees of freedom 
we looked for. Nevertheless, we would like to comment briefly about the emergence of  nonperturbative 
effects in the observables. They will appear  as non-potential terms. At next-to-leading order in the multipole 
expansion they will be encoded into two-point gluonic correlators
which appear in formulas like the following when calculating the energy shift,
\begin{equation}
\sim {1\over T}\int^{T/2}_{-T/2} dt\int^{t}_{-T/2}dt^{\prime}\langle {\bf E}_{i}^a(t)\phi (t,
 t^{\prime})^{adj}_{ab}{\bf E}_{j}^b(t^{\prime})\rangle e^{-i(T/2-t) h_s}{\bf r}^i
 e^{-i(t-t^{\prime}) h_o} {\bf r}^j  e^{-i(t^{\prime}+T/2) h_s}
\label{corHH}
\,,
\end{equation}
where $h_s ={\bf p}^2/m +V_s(r)$ and $h_o={\bf p}^2/m +V_o(r)$.
Notice that the nonperturbative contributions get entangled with the spectrum of the singlet and octet Hamiltonians ($h_s$ and $h_o$) in a non-trivial way.
More information on the gluonic correlator in (\ref{corHH}) cannot 
be obtained up to now from first principle QCD, but rely on lattice simulations \cite{latcor}, 
on QCD vacuum models \cite{DS,BBDV} or on sum rules \cite{Dosch}. 
Still, if we further consider the 
situation $mv^2 \gg \Lambda_{\rm QCD}$ the nonperturbative dynamics can be parameterized into local condensates.
This situation is that one described by Voloshin and Leutwyler \cite{Voloshin}.
It follows from (\ref{corHH}) upon realising that $h_s-h_o \sim mv^2 \gg 
\Lambda_{\rm QCD}$ will make the exponentials oscillate wildly unless $t\sim
t^\prime$, so that a local expansion of the correlator is justified.
On the other hand, if we are in the situation $\Lambda_{\rm QCD} \sim mv^2$, we can not further simplify the 
functional dependence on the nonperturbative dynamics and the observables
depend on non-local condensates. To our knowledge this situation has only been explored in a model dependent context in \cite{mar}. 

\section{pNRQCD: $mv \gg \Lambda_{\rm QCD} \gg mv^2$}
\label{secl2}
In this section we study the situation $mv \gg \Lambda_{\rm QCD} \gg mv^2$. We shall restrict ourselves to pure gluodynamics except in subsection 
\ref{seclig} where we briefly discuss the case of QCD with light fermions. 
Hence, $\Lambda_{\rm QCD}$ here should be understood as a scale of the order
of the mass of the lightest glueball.

In this situation, the pNRQCD Lagrangian of Eq. (\ref{pnrqcd0}), which, we remember, 
has been obtained from NRQCD by simply integrating out the soft scale $mv$, 
does not only describe ultrasoft degrees of freedom ($\sim mv^2$)
but also degrees of freedom associated with the scale $\lQ$ 
which are still dynamical.
If we are only interested in the lower lying states, it is convenient to
 integrate out the scale $\lQ$ in order to 
obtain a suitable effective theory for 
US degrees of freedom only. We will denote this new effective theory by pNRQCD$^{\prime}$.
Since the nonperturbative dynamics is important in this case, it is more difficult to obtain 
model-independent results. Nevertheless, the multipole expansion still holds when matching 
to  
pNRQCD$^{\prime}$. This point is of utmost importance in order 
to control how nonperturbative effects are actually going to appear. If we work at the lowest order in 
the multipole expansion, while the singlet decouples from the octet and the gluons, the octet is still coupled to gluons. 
Let us call gluelumps the adjoint source in the presence of a gluonic field \cite{Michael},\footnote{We will 
see in section \ref{hybrids} that these eigenstates share some 
relation with the energies of gluonic excitations between static quarks in the short-distance limit.} 
$$
H({\bf R},{\bf r},t) \equiv H^a({\bf R},t)O^a({\bf R},{\bf r},t).
$$
We can figure out two situations for the spectrum of these gluelumps 
(leaving aside the perturbative potential).
\begin{itemize}
\item[{\it i)}]{The nonperturbative dynamics creates a gap $\Lambda_{\rm H} \sim \Lambda_{\rm QCD}$ 
in the gluelump spectrum, where $\Lambda_{\rm H}$ is the mass of a generic gluelump.}
\item[{\it ii)}]{The nonperturbative dynamics creates at most a gap $\Lambda_{\rm H} \siml mv^2 \ll \lQ$ 
in the gluelump spectrum.}
\end{itemize}
Situation {\it i)} is the generic one from a Wilson renormalization group point of view 
(assuming naturalness of the coefficients) and hence the most likely to occur. However, situation 
{\it ii)} cannot be ruled out from first principles. We shall briefly study these two situations in the following.

\subsection{$\Lambda_{\rm H} \sim \Lambda_{\rm QCD}$}
As we have just mentioned the pNRQCD Lagrangian of Eq. (\ref{pnrqcd0}) is not the most suitable effective 
theory in order to describe the lower lying states in this situation. In particular, since there is a scale 
($\Lambda_{\rm QCD}$) between the soft scale  $m v$ and the ultrasoft scale $m v^2$, the singlet matching potential 
$V_s$ can no longer be identified with the 
potential which should be used 
in a Schr\"odinger equation. However, the  latter can be obtained by integrating out 
also the scale $\Lambda_{\rm QCD}$, i.e. by constructing pNRQCD$^\prime$. 
Since at scales smaller than $\Lambda_{\rm QCD}$ the nonperturbative effects dominate, it is convenient to have in mind a representation of both pNRQCD 
and pNRQCD$^\prime$ in terms of hadronic fields.
Let us identify the relevant hadronic degrees of freedom of  pNRQCD$^\prime$. Since 
gluelumps and glueballs in pNRQCD have masses 
much larger than $mv^2$ they are integrated out and the only ultrasoft field left is the singlet. 
Let us elaborate on this point using the original fields of pNRQCD. 
The fact that all gluelumps have a gap implies that the octet field develops a
gap and must be integrated out. Hence the only possible remaining fields are
the singlet and ultrasoft gluon fields.
If we assume that the ultrasoft gluon fields have indeed to be kept, terms as
the following should be considered,
\begin{equation}
S^{\dagger}S{\bf r}^i{\bf r}^j\left( c_{EE} {\rm Tr} \{ {\bf E}^i {\bf E}^j\} +
c_{BB} {\rm Tr} \{{\bf B}^i {\bf B}^j\} \right)
\,.
\label{GGg}
\end{equation}  
The term above translates in a hadronic representation to a coupling between quarkonia and glueballs. 
However, for pure gluodynamics the mass of the glueball is believed to be $\sim \lQ$, a scale we have already 
integrated out. Hence we arrive at a contradiction. We conclude that for the case of pure gluodynamics 
no ultrasoft gluon fields have to be included in the pNRQCD$^\prime$ Lagrangian and hence we demonstrate that in this 
situation the quarkonium dynamics reduces exactly to the motion in a
potential. Non-Potential effects do not exist.
Therefore, the pNRQCD$^\prime$ Lagrangian reads
\begin{equation}
{\cal L}_{\rm pNRQCD^\prime} = {\rm Tr} \left\{ {\rm S}^\dagger \left( i\partial_0 - {{\bf p}^2\over m} 
- V^\prime_s(r) + \dots  \right) {\rm S} \right\}, 
\label{pnrqcd1}
\end{equation}
where the dots indicate higher-order potentials in the $1/m$ expansion and the centre-of-mass kinetic 
terms. They are irrelevant here and hence will be neglected. 

$V_s^\prime$ is the potential describing systems characterized by 
$mv \gg \Lambda_{\rm QCD} \gg mv^2$. It has to be obtained by matching pNRQCD with pNRQCD$^\prime$. 
Since $\Lambda_{\rm QCD}$ has been integrated out, the matching is nonperturbative 
and $V_s^\prime$ will contain nonperturbative corrections to the already calculated 
(perturbative) singlet matching potential  of pNRQCD, $V_s(r,\mu) = - C_F \displaystyle{\alpha_{V_s}(r,\mu)\over r}$, 
see Eq. (\ref{newpot}).  These nonperturbative corrections can be computed systematically in the matching procedure 
since the multipole expansion still holds ($mv \gg \Lambda_{\rm QCD}$). In particular at the next-to-leading 
order in the multipole expansion (using the results of section \ref{subsecnlompvs}) we get 
\begin{equation}
V^\prime_s=-C_F {\alpha_{V_s}(r,\mu) \over r } 
-i{g^2 \over N_c}T_F V_A^2(r){r^2 \over d-1} \int_0^\infty \!\! dt 
e^{-it(V_o-V_s)} \langle {\bf E}^a(t) \phi(t,0)^{\rm adj}_{ab}{\bf E}^b(0) \rangle(\mu).
\label{vsprime}
\end{equation}
$\mu$ is the UV cut-off of pNRQCD ($\mu > \Lambda_{\rm QCD}$). 
We notice that Eq. (\ref{vsprime}) is close to that one obtained by Balitsky in \cite{Balitsky}.
Since $V_o-V_s \sim mv^2 \ll \Lambda_{\rm QCD}$ in order to perform correctly the matching 
we have to expand Eq. (\ref{vsprime}) in $V_o-V_s$. In particular, this expansion guarantees that 
only the nonperturbative part (which contains a scale) of the non-local gluon condensate on the right-hand side 
of Eq. (\ref{vsprime}) contributes (in dimensional regularization). Therefore, we get  
\begin{eqnarray}
V^\prime_s&=&-C_F {\alpha_{V_s}(r,\mu) \over r } 
-i{g^2 \over N_c}T_F V_A^2(r){r^2 \over d-1} \int_0^\infty \!\! dt 
\left\{ 1 -it(V_o-V_s) - {1\over 2}t^2(V_o-V_s)^2 \right. \nonumber\\
& & \qquad\qquad\qquad\qquad \left. + i {1\over 6}t^3(V_o-V_s)^3 + \dots \right\}
\langle {\bf E}^a(t) \phi(t,0)^{\rm adj}_{ab}{\bf E}^b(0)\rangle(\mu)^{\rm nonpert.}.
\label{vsprime2}
\end{eqnarray}
Each new term of the expansion is suppressed by an additional factor 
$(V_0-V_s)/\Lambda_{\rm QCD}$.  The leading term cancels the $O(r^2)$ renormalon 
of the perturbative potential (see appendix B), while the term proportional to $(V_0-V_s)^3$ cancels 
the $\ln{\mu}$ dependence of $\alpha_{V_s}(r,\mu)$ (see Eq. (\ref{newpot})). 
In the following subsection we will explicitly show this cancellation 
assuming for the non-local condensate of Eq. (\ref{vsprime2}) a simple model.

Finally note that, although the nonperturbative correction can be
systematically computed in the multipole expansion, the leading $O(r^2)$ 
nonperturbative term could be as important as the perturbative potential once 
the power counting is established, if so, it should be kept exact when solving 
the Schr\"odinger equation.

\subsubsection{A Model with a Massive Gluon Propagator}
In general, the non-local gluon condensate $\langle {\bf E}^a(t) \phi(t,0)^{\rm adj}_{ab}{\bf E}^b(0)\rangle$ 
should  reproduce at high energies (let us say at the scale $\mu$ of the matching between 
NRQCD and pNRQCD) the perturbative expression (\ref{mgeq0}) while at the scale 
of the matching between pNRQCD and pNRQCD$^\prime$ it is dominated by nonperturbative 
physics. A crude assumption (which however satisfies these requests) is that it can be effectively 
described by a massive gluon, where the mass $m_g \simeq \Lambda_{\rm QCD}$ plays the role 
of a hard cut-off at the scale of the matching between pNRQCD 
and pNRQCD$^\prime$. 
Then, we have
\begin{equation}
T_F \langle E^a_i(t) \phi(t,0)^{\rm adj}_{ab}E^b_i(0)\rangle(\mu)^{\rm nonpert.}  = 
C_F C_A \mu^{4-d} \int {d^{d-1}k\over {(2\pi)^{d-1}}}e^{-ik|t|}  
\left({d-2 \over 2} \omega_k + {m_g^2\over 2\omega_k} \right) , 
\label{model} 
\end{equation}
where $w_k\equiv\sqrt{k^2+m_g^2}$. Putting Eq. (\ref{model}) in
Eq. (\ref{vsprime2}) and using (in Euclidean space)
$$
\int { d^d k \over (2\pi)^d} {1 \over (a^2+k^2)^n} = 
{ (a^2)^{{d \over 2}-n} \over (4\pi)^{d \over 2}}{\Gamma(n-d/2) \over \Gamma(n)}, 
$$
we get, after renormalization in the $\overline{MS}$ scheme and neglecting $\alpha_{\rm s}$ corrections in $V_A$,   
\begin{equation}
V^\prime_s=-C_F {\alpha_{V_s}(r,\mu) \over r} 
 +{\alpha_{\rm s} \over 3} C_F r^2 \left\{{m_g^3\over 2} 
+ {3\over 4} (V_o-V_s)^2 m_g - {(V_o-V_s)^3\over\pi} 
\ln {m_g^2\over 4\pi\mu^2}\right\}. 
\label{vsmodel}
\end{equation}
The leading nonperturbative correction in the multipole expansions (which is an expansion in $mv^2/mv$ and  
$m v^2/\Lambda_{\rm QCD}$) to the static potential is in this model of order $r^2\Lambda_{\rm QCD}^3$ 
with the ``right'' positive slope. The correction of order $(V_o-V_s)^3$ cancels the $\ln \mu$ dependence of the 
perturbative potential (see Eq. (\ref{vsus})) so that the expression 
(\ref{vsmodel}) is scale independent (as it should be in a theory, like pNRQCD$^\prime$, free from 
ultrasoft corrections). $V^\prime_s$ contains, therefore, a term like 
$\displaystyle {\alpha_{\rm s}^4 \over r} \ln{rm_g}$ 
which shows that, also in a situation where  potential-like 
nonperturbative corrections exist, it would have been wrong to interpret the re-summed 
static Wilson loop (\`a la Appelquist--Dine--Muzinich) as the singlet static potential. 

Finally we stress that the above results show how, in the physical situation
$mv \gg \Lambda_{\rm QCD} \gg mv^2$, 
nonperturbative contributions cancel the scale dependence of the matching potential $V_s$ in pNRQCD, giving rise 
to a scale independent static potential $V_s^\prime$.  The price to pay is that 
now the static potential is sensitive to the nonperturbative physics. Since we are 
still in the region $r \sim 1/mv < 1/\Lambda_{\rm QCD}$ the computed nonperturbative 
corrections show up as a power series in $r$. 
In principle all powers are allowed. In the model we have used the leading correction 
goes like $r^2$. Since some authors (see for instance \cite{Zakharov}) 
claim the existence of a linear potential at short-distances, it is worthwhile to mention 
that such a kind of correction is in principle not excluded even in this model. 
They may show up at higher order in the multipole expansions (i.e. suppressed 
by powers in $mv^2/mv$ and  $m v^2/\Lambda_{\rm QCD}$).

\subsection{$\Lambda_{\rm H} \siml mv^2 \ll \lQ$}
If some gluelump fields do not develop a gap, they must be retained as low
energy degrees of freedom. Then, the hadronic pNRQCD$^\prime$ Lagrangian would
contain the light gluelump fields in addition to the singlet. 
In terms of the original fields in pNRQCD, pNRQCD$^\prime$ would still contain
the octet in addition to the singlet field. No ultrasoft gluon fields would have to 
be included for the same arguments as in the previous subsection,
and hence non-potential effects would not exist. The singlet and octet matching potentials would receive further contributions from the integration of 
$\lQ$ in a similar way as the singlet matching potential did in the previous subsection. 
Furthermore, at $O(r)$ in the multipole expansion, we would have mixing 
of the light gluelumps and the singlet.
 
It is worth stressing that this situation 
does not correspond to standard potential models. Since we have mixing of different degrees of freedom 
(gluelumps) with what would be the usual quarkonium (the singlet), it rather corresponds to 
a 
coupled channel system in quantum mechanics. 

\subsection{pNRQCD$^\prime$ with light quarks}
\label{seclig}
Let us sketch how the effective theory we have called pNRQCD$^\prime$ is modified 
if light quarks are taken into account, i.e. for real QCD. We first recall at this point that the 
nonperturbative dynamics of pure gluodynamics and that  of  QCD with (more than one) chiral fermions 
has important differences. Whereas the former is believed to have a gap ($\sim \lQ$) in the spectrum 
(being the $0^{++}$ glueball the state with lowest energy), the latter is gapless in the limit of exact 
chiral symmetry since it contains the Goldstone bosons associated to its spontaneous breakdown. 
Still, also for QCD with (more than one) chiral fermions it is believed (and experimentally confirmed) 
that there is a mass gap between the Goldstone bosons and the rest of the spectrum that is 
usually denoted by $\Lambda_{\chi}$, of the order of the $\rho$ mass 
(while it is not clear in this case 
whether the glueball is a stable particle).   

Within this framework, and assuming that $\Lambda_{\chi} \sim \lQ$, we can use $\Lambda_{\chi}$ 
just in the same way as $\lQ$ was used in  pure gluodynamics. 
The analysis below is parallel to the one carried out for the latter case.
Since $mv \gg \lQ \gg mv^2$, 
the multipole 
expansion is valid when matching pNRQCD to pNRQCD$^{\prime}$. At lowest order 
in the multipole expansion, while the singlet decouples from octet and gluons 
(and also light quarks), the octet is still coupled to gluons, and these to
 light quarks. 
Again, we can figure out two situations on the spectrum of the gluelumps:
\begin{itemize}
\item[{\it i)}]{The nonperturbative dynamics creates a gap $\Lambda_{\rm H} \sim \Lambda_{\rm QCD}$ 
in the gluelump spectrum;}
\item[{\it ii)}]{The nonperturbative dynamics creates at most a gap $\Lambda_{\rm H} \siml mv^2 \ll \lQ$ 
in the gluelump spectrum.}
\end{itemize}
Independently of these two situations, 
there will exist extra hadronic ultrasoft degrees of freedom with respect to the case of pure gluodynamics. 
These are the Goldstone bosons associated to the spontaneous chiral symmetry breaking, 
namely pions and kaons. Hence, non-potential effects will exist. This 
important feature will also survive 
in the situation $mv \sim \lQ$.

\medskip

In the case $\Lambda_{\rm H} \sim \lQ$, some results of pure gluodynamics remain 
true. All states with energies of $O(\lQ)$ are integrated out. Now, the 
relevant hadronic degrees of freedom of pNRQCD$^\prime$ are, besides the singlet, the Goldstone bosons. 
Therefore, the hadronic Lagrangian in this case will  have additional terms, besides Eq. (\ref{pnrqcd1}), 
taking care of the interaction of the singlet with the Goldstone bosons. They
enter at $O(r^2)$ in the multipole expansion. 
In terms of the original degrees of freedom (gluons and light quarks), some of
them would read as follows 
$$
 S^{\dagger}S r^2 {\bar q} \gamma^0 q
\,.
$$
Upon hadronisation these terms will give rise to  suitable couplings of Goldstone bosons to quarkonium.

\medskip

In the case $\Lambda_{\rm H} \siml mv^2$, the hadronic degrees of freedom of
pNRQCD$^\prime$ are now, besides the singlet and the light gluelumps, the
Goldstone bosons (again all states with energies of $O(\lQ)$ are integrated
out). Now, the hadronic Lagrangian will have additional terms
taking care of the interaction of the singlet and/or gluelumps with Goldstone
bosons. At $O(r)$ in the multipole expansion
there are no singlet-singlet-Goldstone bosons terms in the Lagrangian, but there can be
 singlet-gluelump-Goldstone bosons, or 
gluelump-gluelump-Goldstone bosons terms. In terms of the original (gluon and light quark) degrees of freedom
the pNRQCD$^\prime$ Lagrangian would look like pNRQCD, that is with the singlet and octet fields, with additional operators involving the light quarks, like
$$
S^\dagger {\rm Tr}({\rm O} T^a)\bar q \gamma^i T^a q {\bf r}^i
$$
which, upon hadronisation would give rise to a singlet-gluelump-Goldstone bosons coupling.

\medskip

Let us finally note that, in principle, the actual values of $\Lambda_{\rm H}$ in pure gluodynamics or in 
real QCD are different since the octet can interact with light quarks through the gluons.

\section{pNRQCD: the static limit}
\label{hybrids}
Although pNRQCD is originally designed to study $Q$-$\bar Q$ systems of large but finite mass, 
it is interesting to study its static limit.
 In
particular, because, in the short-distance limit, it corresponds to the study of the gluelumps defined in the previous section. These, in turn, correspond 
to the gluonic excitations between static quarks in the short-distance limit, for which there is abundant 
nonperturbative data available from lattice simulations. We shall restrict ourselves to the case of pure gluodynamics.

In the static limit there are no space derivatives in the Lagrangian, and hence {\bf r} and {\bf R} are good 
quantum numbers. The spectrum then consists of static energies which depend on {\bf r} 
(translation invariance forbids {\bf R} dependences) and on the only other
scale in this problem, $\lQ$. 
In this section we will discuss some general properties of the short-distance behaviour 
(the only accessible to our analysis) of these static energies. These can be straightforwardly derived in our formalism. 

In the limit $\lQ \ll 1/r$, the spectrum of the theory can be read from  
the Lagrangian (\ref{pnrqcd0}). In particular, the leading order solution corresponds to the zeroth order
of the multipole expansion. At this order the dynamics of the singlet and octet
fields decouple. Hence, the gluonic excitations between static 
quarks in the short-distance limit correspond to the gluelumps. Depending on the  glue operator $H$ and 
its symmetries, the gluelump operator $ O^a H^a$ describes a specific gluonic 
excitation between static quarks and its static energy $V_H$.

In NRQCD (as in pNRQCD) gluonic excitations between static quarks have the same symmetries 
of a diatomic molecule. 
In the centre-of-mass system these correspond to the symmetry group $D_{\infty
  h}$ (substituting the parity generator by CP). 
According to it the mass eigenstates are classified in terms of the angular momentum 
along the quark-antiquark axes ($|L_z| = 0,1,2, \dots$ to which one gives the traditional 
names $\Sigma, \Pi, \Delta, \dots$), CP (even, $g$, or odd, $u$), and the
reflection properties  with respect to 
a plane passing through the quark-antiquark axes (even, $+$, or odd, $-$). 
Only the $\Sigma$ states are not degenerate with respect to the reflection symmetry. 

In pNRQCD at lowest order in the multipole expansion, besides the already 
mentioned symmetries, extra symmetries for the gluonic excitation between static quarks appear.  The glue 
dynamics no longer involves   the relative coordinate $\bf r$. Therefore, 
the glue associated with a gluonic excitation between static quarks acquires a spherical symmetry.
In the centre-of-mass system gluonic excitations between static quarks are, therefore, classified 
according to representations of $O(3) \times$ C, which we summarize by $L$, the angular momentum, CP 
and reflection with respect a plane passing through the quark-antiquark axes\footnote{We could also choose 
P but we prefer this representation for a better comparison with the NRQCD 
quantum numbers.}.
 Since this group is larger than that one of NRQCD, several 
gluonic excitations between static quarks are expected to be approximately degenerate in pNRQCD, 
i.e. in the short-distance limit
$r \ll 1/\lQ$.
We illustrate this point by building up all operators, $H$, up to 
dimension 3 and by classifying them according to their quantum numbers in NRQCD and pNRQCD in Tab. \ref{tab3}. 
In Tab. \ref{tab3} all the operators are intended evaluated in the centre-of-mass coordinates. 
$\Sigma_g^+$ is not displayed since it corresponds to the singlet state. 
The prime indicates excited states of the same quantum numbers. 
The chosen operators for the $\Pi$ and $\Delta$ states are not eigenstates of the reflection operator. 
This is not important since these states are degenerate with respect to this symmetry.
Let us consider, for instance, the $\Pi_g$. The corresponding odd and even states  
with respect the reflection symmetry are given by 
$\left({\bf r}\times\rm{Tr}\{{\bf E} {\rm O}\} \right)_\perp$ and 
$\left({\bf r}\times\rm{Tr}\{{\bf E} {\rm O}\} \right)_\|$ which are the projections 
orthogonal and parallel with respect to the reflection plane respectively.  
The operators ${\bf E}$ and ${\bf D}\times {\bf B}$ (${\bf B}$ and ${\bf D}\times {\bf E}$) 
have the same quantum numbers so by default one would expect them 
to project over the same states but we can go beyond that and see that they are
related by the equations of motion assuring that they will certainly project over the same states. 
From the results of Tab. \ref{tab3} the following degeneracies are expected in
the short-distance limit:
\begin{eqnarray}
&&\Sigma_g^{+\, \prime} \sim \Pi_g\;; \qquad
\Sigma_g^{-} \sim \Pi_g^{\prime} \sim \Delta_g\,; 
\nonumber
\\
&&
\Sigma_u^{-} \sim \Pi_u\;; \qquad
\Sigma_u^{+} \sim \Pi_u^{\prime} \sim \Delta_u \,.
\label{dege}
\end{eqnarray}  
Similar observations have also been made in \cite{Foster}. In pNRQCD they  emerge  in a quite clear 
and straightforward way. Moreover, here we can write explicitly the relevant operators.

\begin{table}[htb]
\makebox[6cm]{\phantom b}
\begin{center}
\begin{tabular}{|c|c|c|}
\hline
Gluelumps       & $~$ & $~$ \\
$ O^a H^a$      & $L=1$ & $L=2$ \\ 
$~$ & $~$ & $~$\\\hline
$\Sigma_g^{+\, \prime}$ & ${\bf r}\cdot{\bf E} \;, 
             {\bf r}\cdot({\bf D}\times {\bf B})$  & $~$  \\\hline
$\Sigma_g^-$ & $~$ & $({\bf r}\cdot {\bf D})({\bf r}\cdot {\bf B}) $ \\\hline
$\Pi_g$ & ${\bf r}\times{\bf E}\;, 
             {\bf r}\times({\bf D}\times {\bf B}) $ & $~$ \\\hline
$\Pi_g^{\prime}$ & $~$ & ${\bf r}\times(({\bf r}\cdot{\bf D}) {\bf B} 
              +{\bf D}({\bf r}\cdot{\bf B}))$  \\\hline
$\Delta_g$ & $~$ & $({\bf r}\times {\bf D})^i({\bf r}\times{\bf B})^j 
             +({\bf r}\times{\bf D})^j({\bf r}\times{\bf B})^i$ \\\hline
\hline
$\Sigma_u^{+}$ & $~$ & $({\bf r}\cdot {\bf D})({\bf r}\cdot {\bf E})$\\\hline
$\Sigma_u^-$ & ${\bf r}\cdot{\bf B} \;, 
             {\bf r}\cdot({\bf D}\times {\bf E})$ & $~$ \\\hline 
$\Pi_u$ & ${\bf r}\times{\bf B}\;, 
             {\bf r}\times({\bf D}\times {\bf E})$ & $~$ \\\hline
$\Pi_u^{\prime}$ & $~$ & ${\bf r}\times(({\bf r}\cdot{\bf D}) {\bf E} 
              +{\bf D}({\bf r}\cdot{\bf E})) $ \\\hline
$\Delta_u$ & $~$ & $({\bf r}\times {\bf D})^i({\bf r}\times{\bf E})^j 
             +({\bf r}\times{\bf D})^j({\bf r}\times{\bf E})^i$ \\\hline
\end{tabular}
\end{center}
\caption{ \it Operators $H$ for the $\Sigma$, $\Pi$ and $\Delta$ gluonic excitations between static quarks in 
pNRQCD up to dimensions 3. The covariant derivative is understood in the 
adjoint representation. ${\bf D}\cdot{\bf B}$ and ${\bf D}\cdot{\bf E}$ do not appear, the first because it 
is identically zero after using the Jacobi identity, while the second gives 
vanishing contributions after using the equations of motion.}
\label{tab3}
\end{table}
 
So far we have just used the symmetries of pNRQCD at lowest order in the 
multipole expansion. In fact we can go beyond that and predict the shape 
of the static energies by actually calculating the correlators 
\begin{equation}
\langle 0|H({\bf R},{\bf r}, T/2) H^{\dagger}
({\bf R}^{\prime},{\bf r}^{\prime}, -T/2)|0 \rangle 
\sim \delta^3({\bf R}-{\bf R}^{\prime})\delta^3({\bf r}-{\bf r}^{\prime})\,e^{-iTV_H(r)}
\end{equation}
for large $T$. At leading order in the multipole expansion we obtain
\begin{equation}
V_H(r) = V_o(r) + {i\over T} 
\ln \langle H^a(T/2) \phi(T/2,-T/2)^{\rm adj}_{ab}H^b(-T/2))\rangle,
\label{vH}
\end{equation}
where the $T\to\infty$ limit is understood. The general structure of the
gluonic correlator is the following (the contribution from the continuum is
included in the dots)
\begin{equation}
\langle H^a(T/2) \phi(T/2,-T/2)^{\rm adj}_{ab}H^b(-T/2)\rangle^{\rm nonpert.} 
\simeq h \, e^{- i \Lambda_H T} + h^\prime\,e^{- i \Lambda_H^\prime T} + 
\dots  
\label{corglHH}
\,.
\end{equation}  
Since we are in the static 
limit, $1/T \ll \Lambda_{\rm QCD} \sim \Lambda_{\rm H} 
< \Lambda_{\rm H}^{\prime} <  \dots$, one can approximate  the right-hand side of Eq. (\ref{vH}) 
for $T\to\infty$ by just keeping the first exponential of Eq. (\ref{corglHH}). 
Then we get at leading order in the multipole expansion 
\begin{equation}
V_H(r) = V_o(r) + \Lambda_H.
\label{vH2}
\end{equation}
Formula (\ref{vH2}) states that at leading order in the multipole 
expansion the short-distance behaviour of the static energies for the gluonic excitations 
between static quarks is described  by the perturbative octet potential plus a nonperturbative constant. 
The constant $\Lambda_H$ depends in general on the particular operator $H$, i.e. on the particular 
gluonic excitation between static quarks. $\Lambda_H$ is the same for operators identifying states 
which are degenerate. 
Notice also that Eq. (\ref{vH2}) can be systematically improved by 
calculating higher orders in the multipole expansion. In particular, 
one can look at how the $O(3)\times$C symmetry is softly broken to $D_{\infty\,h}$ in the short-distance limit.
\begin{figure}[htb]
\makebox[2.5cm]{\phantom b}
\epsfxsize=10truecm \epsfbox{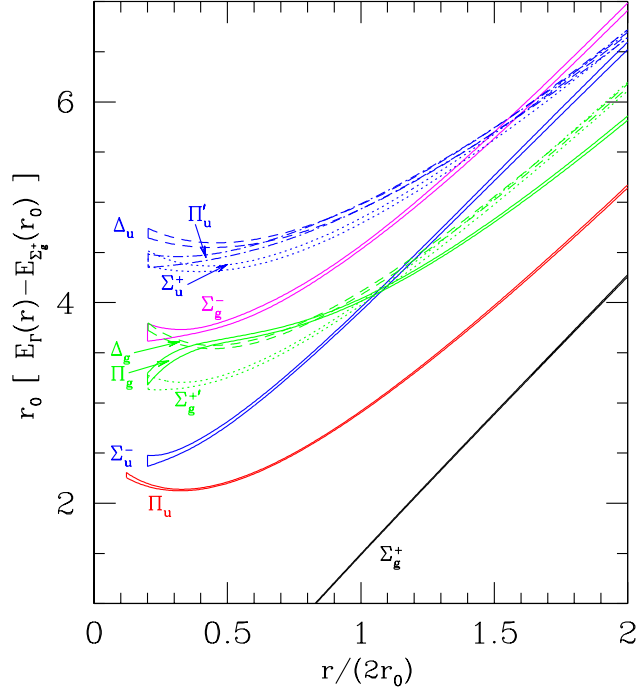}
\vspace{0.2cm}
\caption{ \it Energies for different gluonic excitation between static quarks at distance $r$  
from the quenched lattice measurements of \cite{Morningstar}, $r_0 \simeq 0.5$ fm. 
The picture is taken from \cite{Michael}.}
\label{fighyb}
\end{figure}

Let us compare our results against the best available lattice data (see
Fig. \ref{fighyb} and \cite{Morningstar}). Let us first note that in
our formalism we can trivially disentangle the gluonic
excitations between static quarks 
from the states composed by the singlet plus glueballs. This is no longer so in
lattice simulations where some care should be taken in order to identify these
different states. We will assume in the following that the states measured in
lattice simulations correspond to the gluonic excitations.    
First of all we observe a tendency of all the static energies except $\Pi_g$ to go up at short-distances which may be interpreted as an indication that they want to follow the octet potential shape as required by (\ref{vH2}).   
The $\Pi_u-\Sigma_u^-$ and $\Delta_u-\Pi_u^\prime -\Sigma_u^+$
static energies also show a strong tendency to form a degenerate doublet and triplet respectively at short-distances as 
precisely predicted in (\ref{dege}). However, something strange is
observed for the $g$ static energies. According to (\ref{dege}) a degenerate doublet and triplet should be observed. We can clearly see the doublet $\Pi_g-\Sigma_g^{+\prime}$ but, as mentioned before, the shape of $\Pi_g$ is not compatible with the octet potential. On the other hand $\Delta_g-\Sigma_g^-$ miss 
a state $\Pi_g^{\prime}$ to complete the triplet. We suspect that the plotted
$\Pi_g$ is a superposition of the real $\Pi_g$ in the doublet and the $\Pi_g^{\prime}$ in the triplet. This would explain its peculiar behaviour at small $r$.
We expect that future  
lattice simulations will be able to disentangle these two states and confirm our predictions. It would also be nice to have lattice data at shorter distances so that (\ref{vH2}) can be further confirmed. 
It is interesting to notice that also the hierarchy of the states, 
as displayed in Fig. \ref{fighyb}, is reflected in the dimensionality of the 
operators of table \ref{tab3}. Lower lying states 
are characterized by lower dimensional operators 
if we use a minimal bases with no electric fields. In particular this allows to understand on simple dimensional grounds the highly non-trivial fact that the doublet and triplet static energies lay close together for the $g$ states but quite far apart for the $u$ ones.

The results given above allow us to relate, in the short-distance limit, the behavior of the 
energies for the gluonic excitations between static quarks with the large time behavior 
of some gluonic correlators, in particular with their correlation 
length. It is particularly appealing that we can extract results for 
the exhaustively studied gauge invariant two-point correlator for the gluon field strength tensor:
\begin{equation}
\langle 0|F^a_{\mu\nu}(t)\phi(t,0)^{\rm adj}_{ab}F^b_{\mu\nu}(0)|0 \rangle \,.
\end{equation}
One can parameterize this correlator as a function of two scalar functions:
\begin{equation}
\langle 0|{\bf E}^a(t)\phi(t,0)^{\rm adj}_{ab}{\bf E}^b(0)|0 \rangle \quad
{\rm and}\quad
\langle 0|{\bf B}^a(t)\phi(t,0)^{\rm adj}_{ab}{\bf B}^b(0)|0 \rangle\,,
\end{equation}
with correlation lengths: $T_E=1/\Lambda_{E}$ and $T_B=1/\Lambda_{B}$ respectively. 
From the results of Ref. \cite{Morningstar} displayed in Fig. \ref{fighyb} we can conclude that
\begin{equation}
T_{E} < T_{B}  \label{E>B} \,.
\end{equation}
So far lattice simulations of the gauge invariant two-point correlator for the gluon field strength tensor 
have not reached enough precision to confirm this behavior \cite{latcor}. 
This would be a nice cross-check of lattice simulations. 
On the other hand, recently, a sum rule calculation \cite{Dosch} found evidence 
in favor of Eq. (\ref{E>B}) and also a lattice computation of the 
gluelump masses \cite{Foster} seem to confirm Eq. (\ref{E>B}). 
It would be highly desirable to have more precise lattice data in the short-distance limit in order 
to test our results more quantitatively.

\section{Conclusions}
\label{seccon}
We have presented the matching between NRQCD and pNRQCD at leading order in
$1/m$ and at next-to-leading 
order in the multipole expansion. For the singlet state we have shown that the three 
loop IR divergence of the potential and the two loop IR divergence of the
normalization factor in NRQCD match 
with the corresponding UV divergences of pNRQCD. This is a non-trivial check which confirms that 
pNRQCD is the correct EFT at ultrasoft scales. We have presented the leading log 
contribution to both the  octet and singlet potentials at three loops. Incidentally, 
the two loop octet potential is not available in the literature although it should be 
easily obtained from the intermediate steps of the calculations carried out in
\cite{twoloop}. This would provide an independent test to our proof
that the octet matching potential is free of IR divergences at two loops.
We have worked out the running of the singlet, octet, $V_{A}$ and  $V_{B}$
potentials and the singlet normalization factor   
at leading order in perturbation theory. We have also shown how IR renormalons affecting the static potential
get cancelled in the effective theory.

\medskip

The static limit has been taken throughout because the matching calculation can be done order by order 
in $1/m$ and the static limit is the leading term, and not because we are particularly interested 
in the dynamics of static colour sources. One should bear in mind that pNRQCD is designed to deal with actual
bound state systems made out of quarks of large but finite mass. This is why it is important 
to separate soft from ultrasoft contributions. Our definitions of the potentials are not 
arbitrary but the suitable ones to achieve this goal. Furthermore, the sequence of matching 
calculations QCD $\rightarrow$ NRQCD and NRQCD $\rightarrow$ pNRQCD guarantees that at any desired 
order in $1/m$ and in the multipole expansion, the results we are going to obtain 
(for ultrasoft energies) using pNRQCD are exactly the same as the ones we would obtain using QCD.
What we gain is that calculations in pNRQCD have several important advantages in comparison with direct 
calculations from QCD: (i) each term in the pNRQCD Lagrangian has a size which is simple to 
estimate, (ii) the calculations at leading orders reduce to quantum-mechanical 
calculations, (iii) non-potential effects can be systematically taken into
account, (iv) perturbative and nonperturbative contributions are disentangled
to a large extend.

\medskip

We believe that pNRQCD clarifies several long standing issues. In particular the proper treatment
of the IR divergences in the singlet potential in perturbative QCD. If our primary object of concern 
is the energy between two static quarks, then the IR divergences can be regulated by the 
resummation of Feynman diagrams originally proposed in  \cite{Appelquist}. However, if our 
primary object of concern is to find a static potential which can be used in a Schr\"odinger 
type equation for actual bound state calculations, this is not the right thing to do. The IR divergences can  be simply cut-off, but the cut-off 
dependent potentials have to be used in a framework where ultrasoft gluons (which are also cut-off dependent) 
are properly taken into account, so that in the calculation of any physical quantity the cut-off 
dependence cancels. This is what pNRQCD does for us. We also would like to emphasize that
the fact that nonperturbative (lattice) evaluations of the singlet potential as the energy 
of two static quarks are free from IR divergences does not imply that this is the suitable 
object to be used in a Schr\"odinger type equation, as we have just argued.

\medskip

We have outlined the suitable effective field theories for the ultrasoft degrees of
freedom when $mv \gg \lQ \gg mv^2$, both in pure gluodynamics and in QCD with
light fermions, i.e. real QCD. In pure gluodynamics, we see that
non-potential effects do not exist. Still, different possibilities appear
depending on whether the gluelumps develop a gap of $O(\lQ)$, or of $O(mv^2)$
or smaller. In the first case, only the singlet remains as a dynamical degree
of freedom at the ultrasoft scale, while the integration of the octet field (and gluons)
at next-to-leading order in the multipole expansion introduces  
nonperturbative terms in the potential which are organized in powers
of $\alpha_{\rm s} / r\Lambda_{\rm QCD} \ll 1$ starting with a quadratic
potential. In the second case, some light gluelumps should be kept as 
dynamical degrees of freedom at the scale  $mv^2$. This situation does not
correspond to standard potential models since we have extra degrees of freedom
(gluelumps) besides what it would be the usual quarkonium (the singlet). In
real QCD, new degrees of freedom, the Goldstone bosons associated to the
spontaneous break down of the flavour SU(3) symmetry (pions and kaons) have to
be added at the ultrasoft scale. Therefore, to the features considered for pure
gluodynamics one has to include the fact that, now, non-potential
effects do exist.

\medskip

Some safe statements beyond perturbation theory can be (and have been) made for pNRQCD in the static limit. 
In particular we have studied, in the short-distance limit, what in 
lattice QCD are usually called gluonic excitations of the static quark
potential. 
The symmetries of pNRQCD imply that these states can be classified according to representations 
of $O(3) \times$ C, which are more restrictive than those corresponding to the symmetry 
group of NRQCD with two static sources, namely $D_{\infty h}$. 
This allows us to identify some approximate degeneracies between states. We
have also provided explicit operators for these states. We have
predicted the shape of their static energies and related these with
the correlation length of some gluonic correlators. Note that energies of glue
in the presence of static quark-antiquark pair are commonly studied as a first step in the Born-Oppenheimer treatment of hybrid heavy-quark mesons.

\medskip

We would like to stress that pNRQCD provides a solid framework for the study
of heavy quarkonium systems in a model independent fashion. It may be useful
both for higher order perturbative calculations and for investigating
nonperturbative effects in a systematic way. We would like to briefly comment
below on a few applications that have not been addressed in this paper.

\medskip

The presented results may become relevant in accurate, model-independent,
determinations of the bottom or top mass. The
former is usually determined either by using sum rules \cite{botsumrul,Beneke1} or by direct
analysis of the $\Upsilon (1S)$ mass \cite{Yndurain,Beneke1}, while the latter could be
determined by the study of top pair production near threshold in the Next
Linear Collider \cite{Beneke}. From the perturbative point of view, the
running of the singlet potential is the first step towards the full
calculation of the leading log correction to the next-to-next-to-leading order
results available at present for the above observables. On the
other hand our analysis provides a model-independent framework to estimate and
parameterize nonperturbative effects.

\medskip

Another situation where pNRQCD could
provide useful information is when $\lQ \sim mv$. Most of
the observed charmonium and bottomonium states correspond to this situation, and hence the results presented here cannot be directly applied. 
This situation requires that the matching between NRQCD and pNRQCD is carried out nonperturbatively.
Nevertheless, many of the features observed in this work survive in  a nonperturbative analysis. 
The potential terms can still be obtained in an expansion in $1/m$, 
while the multipole expansion can also be applied for the ultrasoft degrees of
freedom. Moreover, in real QCD the Goldstone bosons (pions and kaons) remain dynamical at
the ultrasoft scale producing non-potential effects. 
In fact, some of the formulas presented here also hold in that situation. 
Work in this direction is in progress \cite{BPSVpre}.

\medskip

{\it Note Added.} After completion of this work Ref. \cite{KP} appeared,
where pNRQCD is applied to the calculation of the US contributions to the 
heavy quarkonium spectrum and to the $Q$-$\bar Q$ production near threshold at
next-to-next-to-next-to-leading order. We have also recently carried out the
complete leading-log next-to-next-to-next-to-leading-order calculation (i.e. the order $m \als^5 \ln \als$, in
the situation $\lQ \ll m\als^2$) of the heavy quarkonium spectrum
\cite{BPSVpre2}. 

\vspace{1cm}

{\bf Acknowledgements.} 
N.B. acknowledges the TMR contract No. ERBFMBICT961714, A.P. the TMR contract No. ERBFMBICT983405, 
J.S. the AEN98-031 (Spain) and 1998SGR 00026 (Catalonia) and A.V. the FWF contract No. P12254.
N.B., J.S. and A.V. acknowledge the program 'Acciones Integradas 1999-2000', project No. 13/99;
thank Gunnar Bali for many interesting and enlightening discussions on the hybrid 
issue and Emilio  Ribeiro of the Technical  University of Lisbon
for hospitality during an earlier stage of this work.  N.B. and A.V. thank the
University of Barcelona for hospitality. 
J.S. thanks the CERN Theory Group, and the Theory Group of the University of Vienna for
hospitality while part of this work was carried out.

\vfill\eject

\appendix

\Appendix{Matching: general remarks}
\label{matgenrem}
In order to clarify the matching procedure adopted in the main part of the
paper, we will study in this appendix the general behaviour of the static Wilson loop, 
\begin{equation}
W_\Box \equiv {\rm P} e^{\displaystyle -i g \oint_{r\times T} dz^\mu A_{\mu}(z)}
\label{wilson0}
\end{equation}
in QED and QCD. In Eq. (\ref{wilson0}) P is the path ordering operator (necessary only for non-commuting 
gauge fields) and the integral is extended on a rectangular path with spatial dimension $r$ and temporal 
dimension $T$. A graphical representation is given in Fig. \ref{wilsonfig}.

\subsection{QED}
\label{secqed}
The Coulomb gauge is implicitly assumed in the discussion below, although, since we are working 
with gauge invariant quantities, the results remain true for any gauge.

Let us consider a world with static fermions and soft photons. 
The static Wilson loop can be computed exactly. If we neglect the 
end-point strings (this is legitimate since, in the Coulomb gauge, the 
$A_0$ component decouples from the transverse photons) the Wilson loop can 
be calculated giving rise exactly to the Coulomb potential $V_{\rm coul}=- \al
/r$. Up to renormalization, the role played by the the soft 
end-point strings is to create states with a definite number of transverse 
{\it soft} photons. The energy difference between states with different number of transverse 
photons is of $O(mv)$ by definition (we are only taking into account soft photons). 
These contributions are exponentially suppressed in the pNRQED power counting, $O(e^{-i/v})$ (since we take 
$V_{\rm coul} \sim mv^2$,  $E_{\rm photon} = k \sim mv$ and  $1/T \sim mv^2$), with respect to the ground state 
contribution (no soft photons in the initial/final state). In fact, they do 
not give contributions to the matching at all. This is easily illustrated if 
we consider the Fourier transform of the Wilson loop with respect to T (taking into account the $\theta(T)$ 
due to the heavy quark propagator). We get
\begin{equation}
\int dT \theta(T) e^{iET}W(\Gamma_0) \simeq {i \over E-V_{\rm coul}+i\epsilon} + 
\int {d^3 {\bf k} \over (2\pi)^3} {i\,Z^{\prime} \over E-V_{\rm coul}-k+i\epsilon} + \dots 
\label{wlstrqed}
\end{equation}
Since we are only interested in $E \sim mv^2$ we are near the first pole and 
all the other contributions can be expanded and do not contribute neither to 
the potential 
nor to the normalization. That is, they do not contribute to the matching.
Specifically, we want to match the expression above to the one obtained in pNRQED
($V_{\rm QED}$ is the potential in pNRQED)
$$
\simeq {i \over E-V_{\rm QED}+i\epsilon}
$$
In this way one trivially gets $V_{\rm QED}=V_{\rm coul}$.

Let us note again that transverse (dynamical) photons do not give any contribution to the 
static potential and that states with transverse soft photons are exponentially suppressed in the Wilson loop. 
$V_{\rm QED}$ is $O(v\simeq \al)$ suppressed with respect to its natural, soft, size due to the $\al$ factor 
coming from the coupling.

\subsection{Perturbative QCD}
Most of what has been said above for QED also holds for perturbative QCD but instead of being 
exact it turns out to be only approximate (up to $O(\alpha_{\rm s})$ corrections).

We again consider a Wilson loop with soft gluons and static quarks in the Coulomb gauge.
We cannot solve the spectrum exactly now. There are interactions  
among the longitudinal and the transverse degrees of freedom. This means that, 
in this case, and unlike in QED, states with a different number of soft transverse gluons 
interact with each other. The analogous of Eq. (\ref{wlstrqed}) reads
\begin{equation}
\int dT \theta(T) e^{iET}W(\Gamma_0) \simeq {i \, Z_s(r)\over E-C_F \als/r(1+\,.\,.\,.)+i\epsilon} + 
\int {d^3 {\bf k} \over (2\pi)^3} {i\, Z^{\prime}_s \over E-k+O(\als)+i\epsilon} + \dots 
\label{wlstrqcd}
\end{equation}
We see that the dependence of 
$\al_{V_s}$ on $\als$ (in the $\overline{MS}$ scheme) is not trivial, $\al_{V_s}=\als(r)+\,.\,.\,.$ 
Moreover, a non-trivial normalization factor, $Z_s(r)$, appears in QCD. The situation in QED 
with light fermions is similar.  

Although we have chosen a straight end-point string in connecting the static quarks another configuration 
could be used. This change of initial/final states produces a different projection on the eigenstates 
of this world of static quarks and soft gluons, but obviously does not change the spectrum. 
As far as we restrict ourself to the matching and to end-point configurations with the same quantum numbers 
as the ground state the change will only affect $Z_s$.

Finally, let us note that the IR cut-off we have introduced in order to 
ensure that only soft degrees of freedom are considered will eventually appear. 
In QCD, as we will see, this already happens for the static Wilson loop. 
In QED, due to the decoupling of the transverse degrees of freedom, the static Wilson loop does not have 
IR problems (although they will appear at higher orders in the $1/m$ expansion).

Summarizing, the role played by the soft scale has been to produce potential
terms and a normalization factor. It should also be noticed that different end-point strings will only change $Z_s$ 
but not the potential.

\begin{figure}[htb]
\makebox[0cm]{\phantom b}
\put(200,10){\epsfxsize=1.5truecm \epsfbox{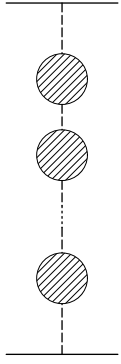}}
\caption{ \it Bubble diagrams contributing to the renormalon of the static potential.}
\label{figren1}
\end{figure}

\Appendix{Renormalons}
\label{secrenormalon}
In this appendix we study the renormalon ambiguities affecting the matching coefficient $V_s$.
For our purposes it is enough to introduce the running of $\als$ at 
one loop within the integral in momentum space (which basically means to 
sum up the leading log of all the bubbles diagrams, see Fig. \ref{figren1}), i.e.  
\begin{equation}
V_s(r)=-4\pi C_F \int {d^3{\bf k} \over (2\pi)^3} {e^{i {\bf k}
\cdot {\bf r}} \over {\bf k}^2}\als({\bf k}^2)\,, \qquad \als({\bf k}^2) \equiv 
{ \als(\mu) \over \displaystyle
1+\beta_0 {\strut {\als(\mu) \over 4 \pi}} \ln{{\bf k}^2 \over \mu^2}}\,.
\label{potren}
\end{equation}

If dimensional regularization is used when matching NRQCD with pNRQCD, 
the integral above runs over all values and Eq. (\ref{potren}) suffers from IR renormalons. 
Let us now see the structure of these singularities. It is sufficient to work with 
a hard UV cut-off such that $1/r \gg \mu \gg \lQ$ and concentrate on scales below 
this cut-off $\mu$ so that the multipole expansion can be applied \cite{Bigi}. We obtain 
the following general expansion for the potential with hard cut-off
\begin{equation}
V_s^{\mu}= C_0 + C_2 r^2 + .\,.\,.
\label{ambpot}
\end{equation}
where
\begin{equation}
C_0=-4\pi C_F \int^{\mu} {d^3{\bf k} \over (2\pi)^3} {1 \over {\bf k}^2} 
\als({\bf k}^2) = - 2{ C_F \als(\mu) \over  \pi} \mu \sum_{n=0}^{\infty}
D_n^{(0)} \left({\beta_0 \als(\mu) \over 4 \pi} \right)^n
\end{equation}
and
\begin{equation}
C_2=2\pi C_F {1 \over 3}\int^{\mu} {d^3{\bf k} \over (2\pi)^3} \als({\bf k}^2) 
= {1 \over 3}{C_F \als(\mu) \over  \pi} \mu^3 \sum_{n=0}^{\infty}
D_n^{(2)} \left({\beta_0 \als(\mu) \over 4 \pi} \right)^n\,.
\end{equation}
The coefficients $D^{(n)}_0$ and $D_n^{(2)}$ are given by 
\begin{equation}
D_n^{(0)}= \int_0^1 dx \left( \ln {1 \over x^2} \right)^n = 2^n\, n!\,,
\qquad
D_n^{(2)} = \int_0^1 dx \, x^2 \left( \ln {1 \over x^2} \right)^n 
= {2^n \over 3^{n+1}}\, n!\,.
\end{equation}
Due to the non-alternating sign of these coefficients, $C_0$ and $C_2$ are not Borel summable and they have 
an ambiguity of $O(\lQ)$ and $O(\lQ^3)$ respectively in their definitions \cite{Agletti}, 
\begin{equation}
\delta C_0 \sim \lQ \,, \qquad \delta C_2 \sim \lQ^3 \,.
\end{equation}

\subsection{Leading order in the multipole expansion}
\begin{figure}[htb]
\makebox[0cm]{\phantom b}
\put(150,10){\epsfxsize=6.5truecm \epsfbox{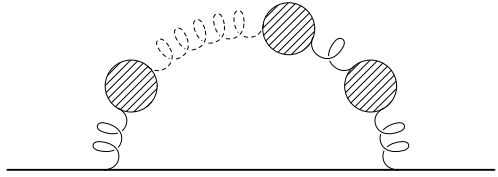}}
\caption{ \it Bubble diagrams contributing to the renormalon of the pole mass.}
\label{figren2}
\end{figure}

The $O(\lQ)$ ambiguity (of IR origin) of $V_s$ cancels against the $O(\lQ)$ ambiguity 
(also of IR origin) of the mass self-energy (see Fig. \ref{figren2}) when computed in the 
pole scheme (which has been implicitly assumed throughout this work)
\cite{thesis}. Therefore, physical observables are free of this renormalon
ambiguity. 

In fact when matching NRQCD with pNRQCD also the heavy quark and antiquark self-energies appear. 
One obtains
\begin{equation}
\label{pmhqet}
2 \Sigma (v \cdot p = 0)  =  -i8\pi C_F \int {d^d k \over (2\pi)^d}
{\als(k^2) \over [ v \cdot k + i\eta]\,[k^2 + i\eta]}
= 4\pi C_F \int{d^{d-1} \, {\bf k} \over (2\pi)^{d-1}}{1 \over{\bf k^2} }\als({\bf k}^2)\,,
\end{equation}
which is zero in dimensional regularization while from a renormalon point of 
view it has an UV and an IR renormalon that cancel each other. Nevertheless, in this respect 
it is nice to see that the full result one obtains when matching NRQCD with pNRQCD, i.e. 
\begin{equation}
2 \Sigma (v \cdot p = 0)  + V_s(r)
\end{equation}
has an UV renormalon while the IR renormalon of each term cancel each other (at lowest order in the 
multipole expansion). This suits the EFT picture where the IR renormalon of twice the pole mass, $O(m)$, 
understood as a matching coefficient,  cancels against the UV renormalon of the next order $O(m^0)$ 
in the $1/m$ expansion.

\subsection{Next-to-Leading order in the multipole expansion}
\begin{figure}[htb]
\makebox[0cm]{\phantom b}
\put(150,10){\epsfxsize=6.5truecm \epsfbox{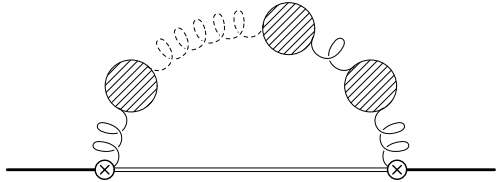}}
\caption{ \it Bubble diagrams contributing in pNRQCD to the renormalon of the singlet self-energy 
         at the next-to-leading order in the multipole expansion.}
\label{figren3}
\end{figure}

The $O(\lQ^3r^2)$ ambiguity of $V_s$ will be absorbed in the effective field theory. It is 
expected to cancel with the renormalon ambiguity of some nonperturbative
effects. Therefore, it is  
different in nature with respect the previous discussed renormalon. The explicit form of these 
nonperturbative effects will depend on the relative size of $\lQ$ with respect the other dynamical 
scales of the problem ($mv^2$, ...) and could be of a potential 
($r^2$ nonperturbative potential if the next relevant scale is $\lQ$, see section \ref{secl2}) 
or non-potential nature (Voloshin-Leutwyler corrections, ... if $\lQ \siml mv^2$, see section \ref{secl1}). 
This is not our problem here but rather to check that the IR renormalon of the matching coefficient 
cancels against the appropriate UV renormalon in pNRQCD (see Fig. \ref{figren3}), 
if pNRQCD has to be a sensible effective field theory. This is, indeed, the case. 

The calculation of the UV renormalon of pNRQCD at next-to-leading order in the multipole expansion 
effectively reduces to the term in Eq. (\ref{vspnrqcdus}) linear in $T$. 
The expression from pNRQCD that should be added to the potential reads
\begin{equation}
-i{g^2 \over N_c}T_F V_A^2(r){r^2 \over d-1} \int_0^\infty \!\! dt 
e^{-it(V_o-V_s)} \langle {\bf E}^a(t) \phi(t,0)^{\rm adj}_{ab}{\bf E}^b(0) \rangle(\mu).
\label{UVpnrqcd}
\end{equation}
Working at the lowest non-trivial order in $\als$ and introducing an IR hard cut-off analogously 
to the procedure followed for the potential (for our purposes here it does not matter 
that the IR behavior is no accurately described), we obtain for Eq. (\ref{UVpnrqcd}) in the UV regime
\begin{equation}
 2\pi C_F {r^2 \over d-1}\int_{\mu} {d^{d-1}{\bf k} \over (2\pi)^{d-1}} \als({\bf k}^2) 
= {r^2 \over d-1}{C_F \als(\mu) \over  \pi} \mu^3 \sum_{n=0}^{\infty}
{\tilde D}_n^{(2)} \left({\beta_0 \als(\mu) \over 4 \pi} \right)^n
\label{UVpnrqcd1} \,.
\end{equation}
where the coefficients are given by
\begin{equation}
{\tilde D}_n^{(2)} = \int_1^\infty dx \,x^{d-2} \left( \ln {1 \over x^2} \right)^n 
\simeq - {2^n \over 3^{n+1}}\, n!
\end{equation}
The non-alternating sign of these coefficients makes Eq. (\ref{UVpnrqcd1}) 
ambiguous by an amount $O(\lQ^3)$, corresponding to the UV renormalon. By closer inspection we can see, as expected, that 
Eq. (\ref{UVpnrqcd1}) cancels against the term $C_2 r^2$ in (\ref{ambpot}), consequently checking the 
cancellation of the renormalon ambiguities up to the considered  order in the multipole expansion.

\vfill\eject

\end{document}